\documentclass[letterpaper,12pt]{article}

\usepackage{bm}
\usepackage{color}
\usepackage{graphicx}
\begin{document}

\title{Hydrodynamic interactions in colloidal ferrofluids: A lattice 
Boltzmann study}

\author{
Eunhye Kim$^1$, 
Kevin Stratford$^2$, 
Philip J. Camp$^3$, and 
Michael E. Cates$^1$ \\
\vspace{1ex} \\
\small
$^1${\it SUPA, School of Physics, The University of Edinburgh}\\
\small
{\it JCMB, The King's Buildings, Mayfield Road, Edinburgh EH9 3JZ, UK}\\
\small
$^2${\it Edinburgh Parallel Computing Centre, The University of Edinburgh}\\
\small
{\it JCMB, The King's Buildings, Mayfield Road, Edinburgh EH9 3JZ, UK}\\
\small
$^3${\it School of Chemistry, The University of Edinburgh} \\
\small
{\it West Mains Road, Edinburgh EH9 3JJ, UK}
}

\maketitle

\begin{abstract} We use lattice Boltzmann simulations, in conjunction with 
Ewald summation methods, to investigate the role of hydrodynamic 
interactions in colloidal suspensions of dipolar particles, such as 
ferrofluids. Our work addresses volume fractions $\phi$ of up to $0.20$ 
and dimensionless dipolar interaction parameters $\lambda$ of up to $8$. 
We compare quantitatively with Brownian dynamics simulations, in which 
many-body hydrodynamic interactions are absent. Monte Carlo data are also 
used to check the accuracy of static properties measured with the lattice 
Boltzmann technique. At equilibrium, hydrodynamic interactions slow down 
both the long-time and the short-time decays of the intermediate 
scattering function $S(q,t)$, for wavevectors close to the peak of the 
static structure factor $S(q)$, by a factor of roughly two. The long-time 
slowing is diminished at high interaction strengths whereas the short-time 
slowing (quantified via the hydrodynamic factor $H(q)$) is less affected 
by the dipolar interactions, despite their strong effect on the pair 
distribution function arising from cluster formation. Cluster formation is 
also studied in transient data following a quench from $\lambda = 0$; 
hydrodynamic interactions slow the formation rate, again by a factor of 
roughly two. \end{abstract}

\section{Introduction} \label{introduction}

The tendency of ferromagnetic colloidal particles to form aggregated 
structures, stabilized by anisotropic dipole-dipole interactions, was 
insightfully discussed by Pierre-Gilles de Gennes and Philip Pincus in 
1970 \cite{deGennes:1970/a}. This tendency is central to the structure 
\cite{Holm:2005/a}, and hence the phase equilibria 
\cite{PJC:2000/a,Tlusty:2000/a,PJC:2007/b} and dynamics 
\cite{PJC:2002/a,PJC:2006/f}, of magnetic colloids in organic solvents 
(ferrofluids). The earliest direct observations of chain-like structures 
were obtained by Hess and Parker using electron microscopy 
\cite{Hess:1966/a} but remarkably the {\em quantitative} experimental 
study of strong pair correlations began only in 2003 with cryogenic 
transmission electron microscopy studies by Philipse and co-workers 
\cite{Butter:2003/a,Butter:2003/b,Klokkenburg:2006/a}. This delay in 
confirming the classic predictions of de Gennes and Pincus partly reflects 
the extreme opacity of most ferrofluids, which precludes both direct 
microscopy of bulk phases and light scattering as methods to elucidate 
structure. Alongside X-ray and neutron scattering, these two methods have 
been central to the widespread progress made in understanding other forms 
of colloidal aggregation since 1970 
\cite{Poon:2002/a,Scheffold:2003/a,Jenkins:2008/a}.

The same experimental difficulties have also made it hard to study 
structural relaxation, which in macroscopically isotropic fluids at 
equilibrium can be quantified by the time-dependent correlator (directly 
accessible in inelastic scattering experiments, where available) 
\cite{Pusey:1991/a,Banchio:2008/a}
\begin{equation} 
S(q,t) = 
\frac{1}{N}\sum_{j,k=1}^N 
\exp{
\left\{ 
 i\bm{q}\cdot\left[ \bm{r}_{j}(t)-\bm{r}_{k}(0) \right] 
\right\}
}
\label{eq:S}
\end{equation}
where $q=|\bm{q}|$. Difficulties with scattering methods mean that many 
relaxation studies on ferrofluids have been limited to strictly $q=0$ 
properties such as frequency dependent bulk magnetic susceptibilities 
\cite{Erne:2003/a,Huke:2004/a} or magnetoviscous and rheological 
properties \cite{Odenbach:2004/a}.

It is increasingly possible for computer simulation methods to fill the 
gaps in our experimental knowledge of structural relaxation in complex 
fluids \cite{Kendon:2001/a,Kremer:2003/a,Cates:2004/a,Stratford:2005/a,% 
Stansell:2006/a,Stratford:2007/a}. However, for ferrofluids, such 
simulations have previously been oversimplified in their neglect of 
hydrodynamic interactions between particles. These interactions are 
mediated by the intervening solvent -- an essentially incompressible, 
Newtonian fluid of viscosity $\eta$ and density $\rho$. Previous 
simulations using molecular dynamics (MD) 
\cite{Wang:2002/a,Wang:2003/a,Huang:2005/a,PJC:2007/c}, Monte Carlo (MC) 
\cite{Weis:1993/a,PJC:2000/c,Levesque:1994/a}, and Brownian dynamics (BD) 
\cite{PJC:2006/f,Meriguet:2004/a,Meriguet:2005/a} of dipolar fluids, while 
each capable of generating the correct equilibrium statistics governed by 
the Boltzmann distribution, are all compromised by either the neglect or 
incomplete treatment of thermal noise and many-body hydrodynamics. In 
particular BD, while capturing the overdamped, diffusive dynamics of an 
isolated particle (with diffusion constant $D_0 = k_BT/6\pi\eta a$, $a$ 
being the particle radius) fails to correlate the Brownian motion of two 
or more particles in the correct manner. As a result, $S(q,t)$ will have 
the wrong time dependence.

In this work, we present simulation results for dipolar colloids 
generated using the lattice Boltzmann (LB) method. This approach treats 
full hydrodynamic interactions, at least for the colloids simulated 
here, which have a sufficiently repulsive soft-core potential to ensure 
that particles do not approach one another too closely. (The presence of 
such a potential avoids large hydrodynamic lubrication forces in thin 
fluid films between two solid particles in close contact, whose 
treatment within LB is possible, but costly \cite{Nguyen:2002/a}.) 
Alongside LB, other methods that fully treat hydrodynamics include force 
methods \cite{Ladd:1990/a}, Stokesian dynamics (SD) and accelerated 
Stokesian dynamics (ASD) \cite{Phung:1996/a,Banchio:2003/a}, and 
stochastic rotation dynamics \cite{Malevanets:1999/a}. (Dissipative 
particle dynamics does so also, but without proper control of noise 
terms as required here \cite{Groot:1997/a}.) Few of these methods have 
yet been used to treat systems with long-range interactions, although an 
SD method has been applied to ferrofluids in shear flow 
\cite{Satoh:1998/a,Satoh:1999/a}, and ASD was recently used to address 
charge-stabilized colloids at low ionic strength \cite{Banchio:2008/a}. 
We are aware of no {\em systematic} application of these methods to the 
equilibrium dynamics of dipolar fluids in three dimensions.

This paper is organized as follows. In section \ref{methods} we outline 
the numerical methodology, and in section \ref{statics} present results 
for equilibrium structure as calculated by BD, LB, and MC methods. (These 
should be, and nearly are, identical.) Then in section \ref{dynamics} we 
present results for $S(q,t)$ and its orientational analogs, focusing on a 
like-for-like comparison between BD (no hydrodynamics) and LB (full 
hydrodynamics). In section \ref{kinetics} we present an analysis of 
transient behavior, addressing the time evolution of the static structure 
including an analysis of cluster statistics. Finally in section 
\ref{conclusions} we give our conclusions and discuss prospects for future 
work.

\section{Simulation methods} \label{methods}

We use the lattice Boltzmann method for a fluid incorporating spherical 
solid particles \cite{Nguyen:2002/a,Stratford:2008/a}. In this method, the 
density, momentum, and stress in the fluid are associated with various 
moments of a kinetic distribution function $f(\bm{c}_i,\bm{r})$, defined 
at each site $\bm{r}$ on a 3D lattice and acting on a space of 19 discrete 
velocities $\bm{c}_i$ that connect neighboring sites in one timestep 
(including the null velocity). Setting the lattice parameter, timestep, 
and mean fluid density $\rho = \langle \sum_{i} f \rangle$ all to unity 
defines a set of LB units that we use in results quoted below. (Of course, 
many physical quantities can be expressed in dimensionless forms, from 
which these units cancel out.)

Each of our spherical colloidal particles has a hard-core radius $a = 2.3$ 
and resides off-lattice; its surface then cuts the lattice bonds at a set 
of links at which fluid and particle interact by momentum transfer. This 
transfer is achieved by a `bounce-back on links' algorithm 
\cite{Nguyen:2002/a}. The force and torque on a colloidal particle are 
found by summing contributions across the boundary links; particle 
velocities and angular velocities are then updated via a standard 
molecular dynamics routine \cite{Nguyen:2002/a,Stratford:2008/a}. The 
thermal noise, responsible for colloidal Brownian motion, is generated 
entirely within the fluid and is fully included in the description of the 
fluid momentum \cite{Ladd:1994/a}, using a method reported previously 
\cite{Adhikari:2005/a}. Momentum transport then causes the random forces 
(and torques) felt by different particles to become correlated, in accord 
with the fluctuation-dissipation theorem which relates random forces to 
the matrix of particle mobilities $M_{ij\alpha\beta}$. This matrix obeys 
$M_{ij\alpha\beta} = \partial {v}_{\alpha,i}/\partial {f}_{\beta,j}$ where 
$\bm{v}_i$ is the velocity of particle $i$ in response to a force 
$\bm{f}_j$ on particle $j$, and $\alpha,\beta$ are cartesian indices. In 
the presence of hydrodynamic interactions, $M_{ij\alpha\beta}$ is not 
diagonal in particle indices, but has long-range correlations. In contrast 
to SD-based methods, LB avoids explicit computation of 
$M_{ij\alpha\beta}$.

Each colloidal particle has the same mass $m = \rho v_{0}$ as the 
nominal volume of fluid that it displaces; $v_{0} = 4\pi a^3 / 3$ is the 
volume of a colloidal particle. The fluid viscosity is set to $\eta = 
0.025$ and the temperature to $k_{B}T = 5\times 10^{-5}$; these choices 
represent the best compromise we have found between numerical accuracy 
and efficiency for the systems under study here. (For a discussion of 
parameter optimization in LB see \cite{Cates:2004/a,Stratford:2005/a}.) 
Based on these parameters, the particle velocity relaxation time is 
$\tau_{v} \equiv m/6\pi\eta a \simeq 47$, and the time scale for fluid 
momentum to equilibrate around a particle is $\tau_{\eta} \equiv 
a^{2}\rho/\eta \simeq 212$. The single-particle diffusivity is $D_0 = 
k_{B}T / 6\pi\eta a \simeq 4.61 \times 10^{-5}$, giving a diffusive 
timescale of $\tau_D \equiv a^2 / D_0 \simeq 1.15 \times 10^{5}$. This 
offers a reasonably wide domain in which to study short time diffusion, 
even in the case of strongly interacting particles (as here) where the 
short-time diffusion time window closes off at $t \ge 
\tau_{D}(\bar{h}/a)^2$ with $\bar{h}$ a typical surface-to-surface 
separation between adjacent particles. Our choices for other (reduced) 
simulation parameters -- to be defined below -- correspond to typical 
values for real ferrofluids. Time-dependent results are expressed in 
units of $\tau_{D}$, which means that for $t \gg \tau_{v},\tau_{\eta}$, 
they will be directly relevant to almost any real ferrofluid system.

To avoid computationally expensive lubrication contacts between particles 
(as mentioned in section \ref{introduction}) we introduce a short-range, 
soft-core repulsion $U^{sc}$ acting at separations beyond the hard-core 
radius $a=2.3$. (The latter coincides with the hydrodynamic radius 
\cite{Nguyen:2002/a,Stratford:2008/a}.) We tested various options for 
$U^{sc}$ and found that a satisfactory choice is
\begin{displaymath}
U^{sc}(h) = 
\left\{
\begin{array}{ll}
 U_{0}(h)-U_{0}(h_{c})-(h-h_{c})dU_{0}/dh|_{h=h_{c}} & 0 < h \leq h_{c} \\
 0                                                   & h > h_{c}
\end{array}
\right.
\end{displaymath}
where $U_{0}(h) = k_{B}Ta/h$, $h=r-2a$ is the surface-to-surface 
separation for spheres whose centers are $r$ apart, and $h_{c}$ is a 
short-range cutoff. We set the cutoff separation $h_{c} = 1.2$, which in 
terms of the hard-core diameter is $h_{c}/2a \simeq 0.26$. This 
comprises a truncated and shifted inverse power law potential; note that 
the interaction force remains divergent at contact ($h=0$). Although 
structural and magnetic properties of ferrofluids are known to be quite 
insensitive to the choice of short-ranged repulsive potential 
\cite{PJC:2007/c}, our choice could qualitatively describe the effects 
of a screened electrostatic repulsion. The case of steric stabilisation 
is more complex since the polymer layer will have, in addition, a direct 
effect on the hydrodynamic forces.

The introduction of $U^{sc}$ reduces discretization errors in the noise 
forces which become acute when fluid nodes are excluded from the space 
between particles. (Since absence of fluid entails absence of noise, 
particles that are too close to one another to have fluid nodes between 
them effectively feel a reduced temperature.) In fact this issue, rather 
than avoidance of lubrication contacts {\it per se}, prevents efficient 
use of a shorter-range soft-core potential than the one we have chosen; 
the effect of using such a potential is to exaggerate the peak in the pair 
distribution function $g(r)$ for particles at close contact (as though 
particles become colder at close separations). Even with our choice of 
$U^{sc}$, this deviation remains visible in the data of section 
\ref{statics} for the largest dipole strength used, but this is considered 
acceptable. Indeed, in the current state of the art for LB, errors of 
several percent from this and other sources such as shape discretization 
remain unavoidable. To reduce these further is straightforward in 
principle: one simply increases $a$. However, for a fixed number of 
particles at a given volume fraction $\phi$, the system volume must then 
be increased as $a^3$, with a further $a^2$ increase in the run time to 
reach $\tau_D$. Thus a factor 2 increase in $a$, as would be required to 
give a worthwhile reduction in discretization errors, entails a 32-fold 
increase in computational resource.

The total colloid-colloid pair potential in our simulations is the sum of 
short-range soft-core ($sc$) and long-range dipolar ($d$) contributions:
\begin{equation} 
U_{ij} = U^{sc}(h_{ij}) 
       + U^{d}(\bm{r}_{ij},\hat{\bm{s}}_i, \hat{\bm{s}}_j).
\label{totalpair}
\end{equation}
Here, $\bm{r}_{ij}$ is the center-center separation vector for particles 
$i$ and $j$, $\hat{\bm{s}}_{i}$ denotes a unit vector pointing along the 
dipole of particle $i$, and $h_{ij} = r_{ij} - 2a$ where $r_{ij} = 
|\bm{r}_{ij}|$. The dipole-dipole interaction potential is written
\begin{equation}
U^{d}(\bm{r}_{ij},\hat{\bm{s}}_i, \hat{\bm{s}}_j) = 
8\lambda k_BTa^3
\left[
 \frac{(\hat{\bm{s}}_i \cdot \hat{\bm{s}}_j) -
       3(\hat{\bm{s}}_i \cdot \hat{\bm{r}}_{ij})
        (\hat{\bm{s}}_j \cdot \hat{\bm{r}}_{ij})}
      {r_{ij}^3}
\right]
\label{dipoleU}
\end{equation}
where $\hat{\bm{r}}_{ij} = \bm{r}_{ij} / r_{ij}$ is a unit vector pointing 
along the center-center separation vector. The interaction parameter 
$\lambda$ is a dimensionless, dipolar coupling constant defined such that 
two colloids at hard core contact ($r=2a$), with dipoles mutually aligned 
and parallel to $\bm{r}$, have $U^{d} = -2\lambda k_{B}T$. This 
`nose-to-tail' parallel conformation is of lowest energy; the other 
minimum-energy conformation is `side-by-side' antiparallel, for which $U^d 
= -\lambda k_BT$ (with corresponding energy maxima on reversing the 
direction of one dipole). In experimental ferrofluids $\lambda$ values up 
to $\simeq 4$ are readily available; much larger ones can be achieved but 
are relatively unusual 
\cite{Butter:2003/a,Butter:2003/b,Klokkenburg:2006/a}.

All of our simulations were performed in a cubic box of side $L$ with 
periodic boundary conditions for both fluid and particles. A standard 
Ewald summation technique was deployed to handle the long-range aspect 
of the dipolar interactions \cite{deLeeuw:1980/a,Allen:1987/a}. The 
Ewald sum computes dipolar forces directly in real space for particle 
pairs with separation $r_{ij} < r_c$, a cutoff distance, and deals with 
the remainder in reciprocal space. To allow a parallel implementation 
using domain decomposition, which is important given the relatively high 
computational requirements associated with the LB fluid 
\cite{Stratford:2008/a}, we chose $r_c = 16$ in all cases.  This is 
small enough that some parallelism is possible, but not so small that 
the number of terms required in the reciprocal space sum for acceptable 
accuracy becomes unwieldy. A convergence factor \cite{Allen:1987/a} of 
$\kappa = 5/2r_c = 0.15625$ was chosen, with wavevectors 
$\bm{k}=(2\pi/L)\bm{n}$ with $|\bm{n}| \leq 8$ and $16$ for $L = 64$ and 
$128$, respectively. These parameters were optimised using established 
methods \cite{Fincham:1993/a}. Finally, following normal practice, the 
Ewald sum boundary condition at infinity was chosen to be ``conducting", 
i.e., the dipolar (magnetic) susceptibility of the surroundings is 
infinite. This removes a zero-wavevector, bulk-magnetization term from 
the Hamiltonian which, in the opposite extreme of vacuum boundary 
conditions, can lead to unphysically small polarised domains and hence 
slow down simulation convergence \cite{Wei:1992/a}.

For comparison with our LB results, a BD algorithm was set up within the 
colloidal MD module of our LB code, deploying standard techniques to 
generate the required Langevin dynamics with independent noise acting 
directly on each particle. The inertia of the particles is retained but 
the many-body hydrodynamics are replaced by a Stokes drag that is 
independent of the location of other particles. By setting exactly the 
same value for $\tau_D$, we thus create an algorithm which differs from LB 
solely in the omission of many-body hydrodynamics.

To further validate both codes, and to monitor the achievement of 
Boltzmann equilibrium, canonical Monte Carlo simulations were performed in 
a cubic simulation cell with periodic boundary conditions applied 
\cite{Allen:1987/a}. The long-range dipolar interactions were handled 
using the Ewald summation with conducting boundary conditions, a 
convergence factor $\kappa L = 5.6$, and wavevectors 
$\bm{k}=(2\pi/L)\bm{n}$ with $|\bm{n}| \leq 6$. The maximum translational 
and orientational displacement parameters were adjusted independently to 
give acceptance rates of approximately $20\%$ and $50\%$, respectively; it 
is efficient to employ low acceptance rates for translations of particles 
with hard cores, due to the possibility of rapidly identifying overlaps. 
For each state point considered, we performed equilibration runs of $2 
\times 10^{5}$ MC cycles, where one MC cycle consisted of, on average, one 
attempted translation or rotation per particle. Production runs consisted 
of $5 \times 10^{5}$ MC cycles.

Our LB work was performed primarily on lattices of size $V=64^3$ or 
$V=128^3$. The colloid volume fraction is given by $\phi = Nv_{0} / V$, 
where $N$ is the number of colloids. For the larger system size at a 
volume fraction $\phi=0.10$, a run of $\sim 10^6$ timesteps required 56 
hours on 64 cores of a cluster of 3GHz Intel dual-core processors. For 
resource reasons most of our results on fully equilibrated samples ($t\ge 
2$-$3\times 10^6 \simeq 25\tau_D$ at $\lambda = 8$) concern the smaller 
$V$. BD and MC runtime requirements were modest in comparison.

\section{Static structure} \label{statics}

In this section we confirm that our LB algorithm generates, to acceptable 
accuracy, the Boltzmann distribution for thermal equilibrium properties, 
as does our BD code, and that both are in agreement with MC data. This is 
of course necessary if the dynamical data in subsequent sections are to be 
trusted. Such agreement is not automatic but is in fact a very demanding 
test of the ability of our LB algorithm to generate correlated noise 
forces satisfying the fluctuation dissipation condition. In particular, 
without adopting the methods of \cite{Adhikari:2005/a} (in which noise 
terms are applied to not only the hydrodynamic but also the local modes of 
the fluid degrees of freedom) we would not be confident of achieving such 
agreement. Even with these methods, LB parameter values must be carefully 
chosen to maintain acceptable performance. We have already mentioned that 
correct treatment of noise was the limiting factor in our choice of 
$U^{sc}$; it also prevents us using a much larger $k_BT$ value (which sets 
the intrinsic noise level in the LB fluid) and/or a much smaller viscosity 
$\eta$. Either step could in principle drastically reduce the run time 
required to reach the basic timescale $\tau_D$ for colloidal diffusion. 
Thus the control of discretization errors within the noise sector 
currently remains the primary efficiency bottleneck in the application of 
large-scale LB simulations to colloidal diffusion.

\subsection{Energy equilibration}

Table \ref{tab:energy} shows time-averaged energy data for various 
simulation runs. Those for $V = 64^3$ have fully equilibrated (run times 
$> 30 \tau_D$) as judged by convergence of the energy parameters to their 
long-term averages. The results indicate excellent agreement between LB 
and the other methods at $\lambda = 0$ and $4$, and adequate agreement 
(errors of less than 5\%) at $\lambda = 8$. Some runs with the larger 
system size, $V = 128^3$, are also reported in the Table. These have run 
times $\sim 7\tau_D$ and while their energies appear to have saturated, 
those at $\lambda = 8$ are showing continued structural evolution by other 
measures (such as cluster statistics; see section \ref{kinetics}). 
Accordingly, the reported energy discrepancies for these runs may include 
systematic errors arising from incomplete structural equilibration, and 
should not be taken as a guide to the relative accuracy of the LB 
algorithm.

%%% Table 1 %%%
\begin{table}[t!]
\centering
\begin{tabular}{llrrrcc}
\hline\hline
$\lambda$ & $\phi$ & $N$ & $V$ & Method & $U^{d}/Nk_BT$ & $U^{sc}/Nk_BT$ \\ \hline
0 & 0.10 &  529 &     $64^3$ & LB & -                     & $0.08671 \pm 0.00028$ \\
  &      &  529 &     $64^3$ & BD & -                     & $0.0870  \pm 0.0004$  \\ 
  &      &  529 & $22156a^3$ & MC & -                     & $0.08757 \pm 0.0001$  \\ \hline
4 & 0.10 &  529 &     $64^3$ & LB &  $-2.929  \pm 0.003$  & $0.2923  \pm 0.0006$  \\
  &      &  529 &     $64^3$ & BD &  $-2.964  \pm 0.002$  & $0.2935  \pm 0.0006$  \\ 
  &      &  529 & $22156a^3$ & MC &  $-2.8830 \pm 0.0008$ & $0.2850  \pm 0.0002$  \\ \hline
8 & 0.10 &  529 &     $64^3$ & LB & $-11.811  \pm 0.002$  & $1.1692  \pm 0.0007$  \\
  &      &  529 &     $64^3$ & BD & $-11.609  \pm 0.002$  & $1.1253  \pm 0.0007$  \\
  &      &  529 & $22156a^3$ & MC & $-11.565  \pm 0.003$  & $1.1196  \pm 0.0006$  \\ \hline
4 & 0.20 & 8239 &    $128^3$ & LB &  $-3.966  \pm 0.001$  & $0.5140  \pm 0.0004$  \\
  &      & 8239 &    $128^3$ & BD &  $-3.902  \pm 0.001$  & $0.4970  \pm 0.0006$  \\
  &      &  529 & $11079a^3$ & MC & $-4.1895  \pm 0.0008$ & $0.4534  \pm 0.0002$  \\ \hline
8 & 0.20 & 8239 &    $128^3$ & LB & $-11.833  \pm 0.003$  & $1.233   \pm 0.001$   \\
  &      & 8239 &    $128^3$ & BD & $-11.646  \pm 0.003$  & $1.188   \pm 0.002$   \\
  &      &  529 & $11079a^3$ & MC & $-11.677  \pm 0.003$  & $1.1925  \pm 0.0006$  \\ \hline\hline
\end{tabular}
\caption{\label{tab:energy} Energy equilibration data for LB, BD, and MC 
simulation runs. The quoted statistical errors are estimated on the basis 
of one standard deviation. $\lambda$ is the dipolar coupling constant and 
$\phi = Nv_{0}/V$ is the volume fraction where $N$ is the number of 
colloids and $v_{0} = 4\pi a^3 / 3$ is the volume of one colloid. The 
system volumes are reported in lattice units for LB and BD runs, while the 
MC volumes are reported in units of the hard-core radius $a$ (equal to 
$2.3$ in lattice units).}
\end{table}

\subsection{Radial distribution functions (RDFs)}

\begin{figure}[t!]
\centering
\begin{tabular}{cc}
\includegraphics[width=2in]{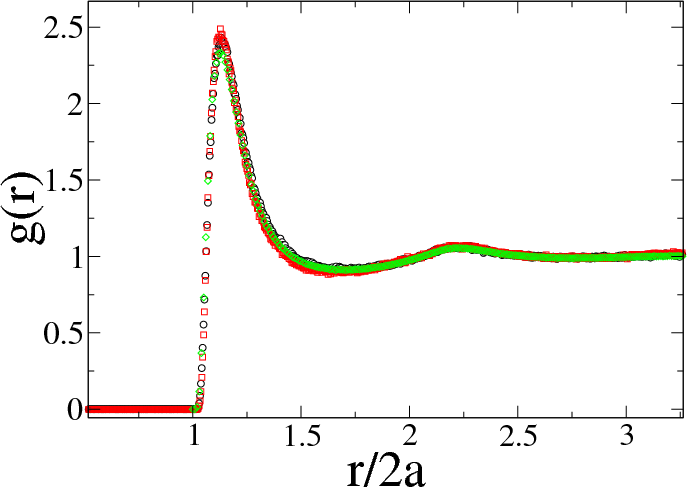} &
\includegraphics[width=2in]{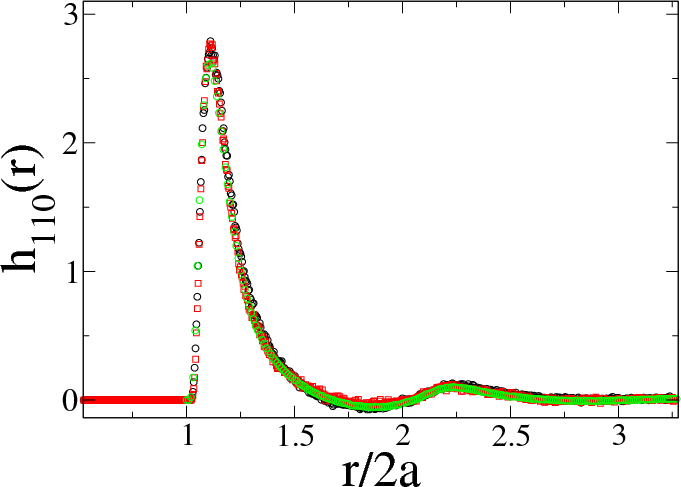} \\
(a) $g(r)$ &
(b) $h_{110}(r)$ \\
\includegraphics[width=2in]{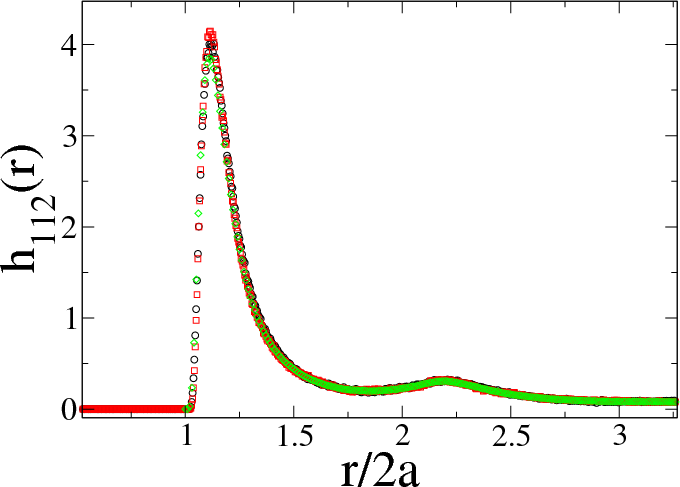} &
\includegraphics[width=2in]{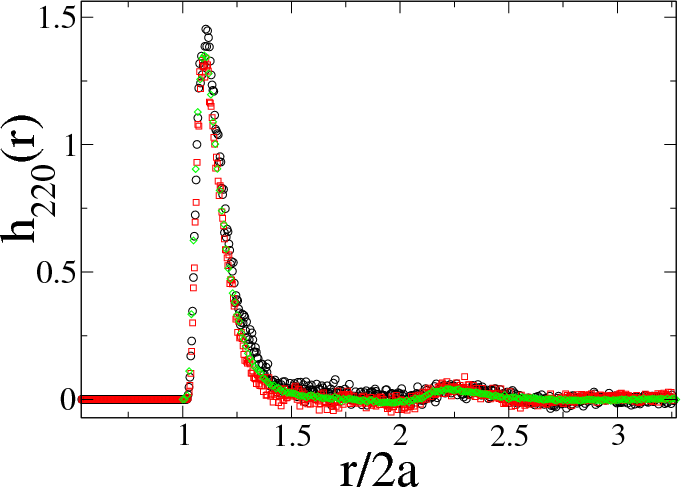} \\
(c) $h_{112}(r)$ &
(d) $h_{220}(r)$
\end{tabular}
\caption{\label{fig:one}Radial Distribution functions (a) $g(r)$, (b) 
$h_{110}(r)$, (c) $h_{112}(r)$, and (d) $h_{220}(r)$ for $\lambda=4$ and 
$\phi=0.10$: (black circles) BD; (red squares) LB; (green diamonds) MC.}
\end{figure}

\begin{figure}[t!]
\centering
\begin{tabular}{cc}
\includegraphics[width=2in]{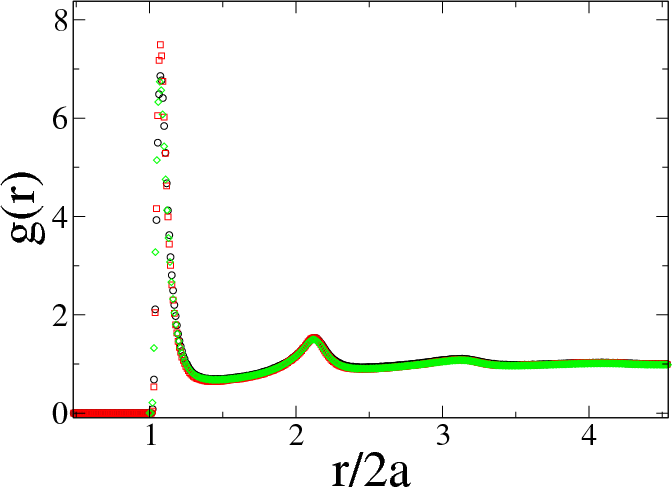} &
\includegraphics[width=2in]{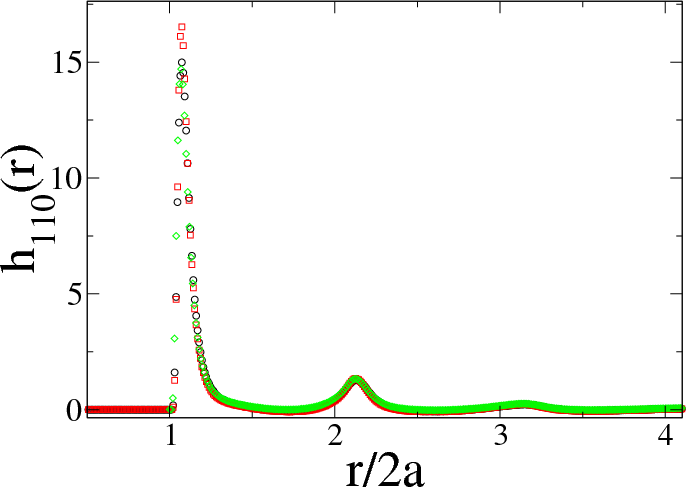} \\
(a) $g(r)$ &
(b) $h_{110}(r)$ \\
\includegraphics[width=2in]{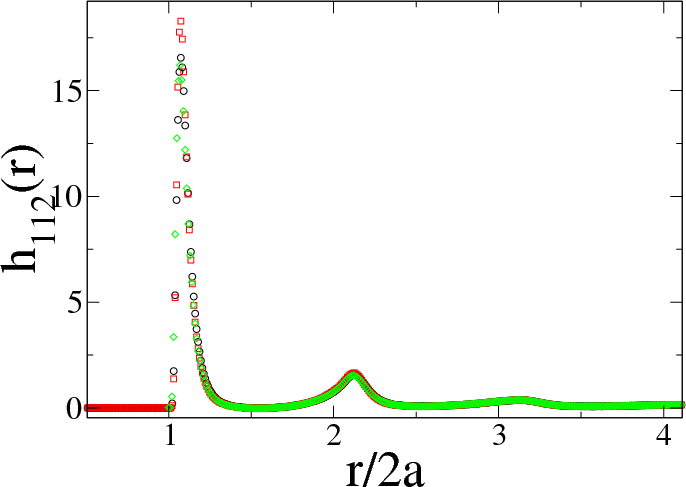} &
\includegraphics[width=2in]{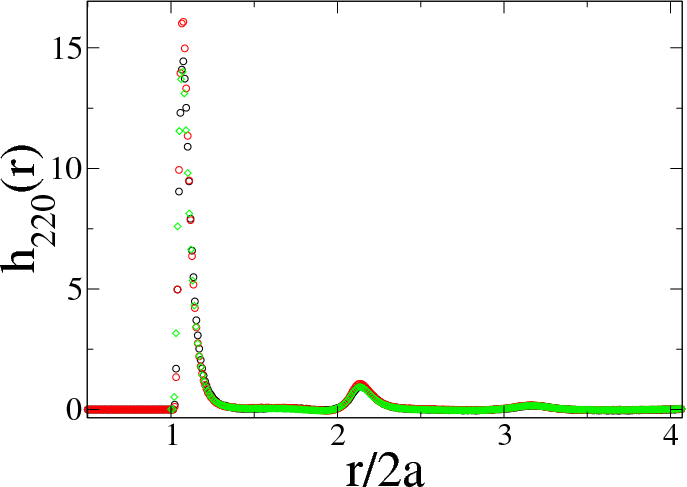} \\
(c) $h_{112}(r)$ &
(d) $h_{220}(r)$
\end{tabular}
\caption{\label{fig:two} Radial distribution functions (a) $g(r)$, (b) 
$h_{110}(r)$, (c) $h_{112}(r)$, and (d) $h_{220}(r)$ for $\lambda=8$ and 
$\phi=0.10$: (black circles) BD; (red squares) LB; (green diamonds) MC.}
\end{figure}

In Figures \ref{fig:one} and \ref{fig:two} we plot the radial distribution 
function $g(r)$, and the projections of the `molecular' pair distribution 
function onto rotational invariants \cite{Hansen:1986/a,Weis:1993/b}, as 
measured in various LB runs:
\begin{eqnarray}
g(r) &=& 
     \frac{V}{2\pi r^2 N^2}
     \left\langle  
      \sum_{i<j} 
      \delta{(r-r_{ij})}
     \right\rangle \\
h_{110}(r) &=& 
           \frac{3V}{2\pi r^2 N^2} 
           \left\langle 
            \sum_{i<j} 
            \delta{(r-r_{ij})}
            (\hat{\bm{s}}_i \cdot \hat{\bm{s}}_j) 
           \right\rangle \\
h_{112}(r) &=& 
           \frac{2}{3} \frac{V}{2\pi r^2 N^2}
           \left\langle 
            \sum_{i<j} 
            \delta{(r-r_{ij})}
            [3(\hat{\bm{s}}_i \cdot \hat{\bm{r}}_{ij})
              (\hat{\bm{s}}_j \cdot \hat{\bm{r}}_{ij})
             -(\hat{\bm{s}}_i \cdot \hat{\bm{s}}_j)] 
           \right\rangle \\
h_{220}(r) &=& 
           \frac{5}{2} \frac{V}{2\pi r^2 N^2}
           \left\langle 
            \sum_{i<j} 
            \delta{(r-r_{ij})}
            [3(\hat{\bm{s}}_i \cdot \hat{\bm{s}}_j )^2 - 1] 
           \right\rangle.
\end{eqnarray}
Note that all the runs reported have $\phi = 0.10$. We do not report 
equilibrium structural data for $\phi = 0.20$ here because, even though 
the energy can be equilibrated for $V = 64^3$, the modest number of 
particles combined with a very long autocorrelation time means that good 
statistics cannot be gathered for structural quantities even with millions 
of timesteps. For $V = 128^3$ the statistics are better due to larger $N$, 
but the maximum available runtime ($\sim 10^6$) is not long enough to 
guarantee equilibration as discussed above.

Each of the above RDFs could equally be plotted in Fourier space (with 
$g(r)$ then transforming into the static structure factor $S(q)$), but 
the real space versions offer the more sensitive tests of equilibration. 
This is because, for the reasons already discussed, any errors are 
likely to occur in a localized range of $r$ at close contact ($r\simeq 
2a$). We find excellent agreement between LB, BD, and MC for $g(r)$ at 
$\lambda = 0$ (data not shown). Figure \ref{fig:one} shows adequate 
agreement between all methods at $\lambda = 4$ although BD and LB both 
show slight discrepancies from the MC data (which should be the most 
accurate) in the neighborhood of the first peak for each correlator. (A 
slight discrepancy for LB is also detectable near the first minimum of 
$h_{110}(r)$.)

%%% Figure 1 %%%

For $\lambda = 8$ -- Figure \ref{fig:two} -- there is a clear discrepancy 
between LB and the other two methods, overestimating by 10\% or so the 
height of the first peak in all the RDFs. This error is consistent in sign 
and magnitude with the energy discrepancies in Table \ref{tab:energy}, and 
with the specific type of noise discretization error reported in section 
\ref{methods}. Given the other sources of error in LB \cite{Cates:2004/a}, 
we consider it acceptable.

%%% Figure 2 %%%

\begin{figure}[b!]
\centering
\begin{tabular}{cc}
\includegraphics[width=2.75in]{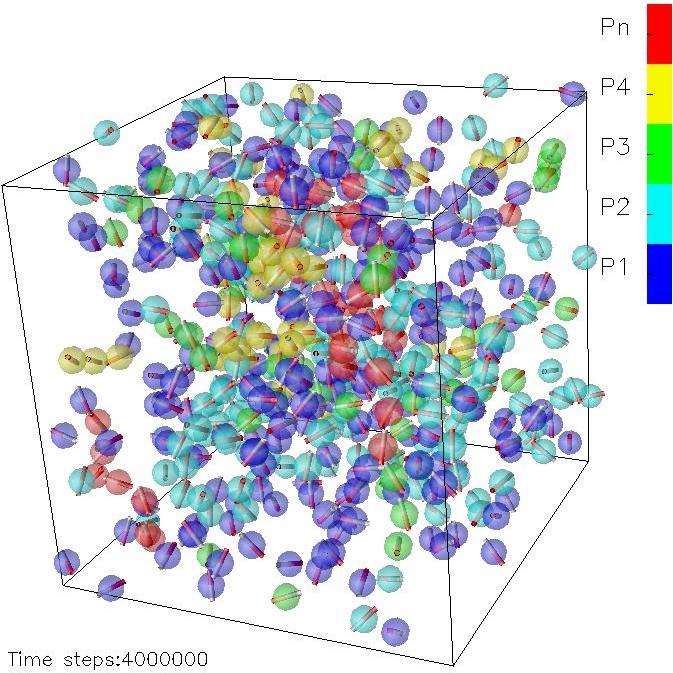} &
\includegraphics[width=2.75in]{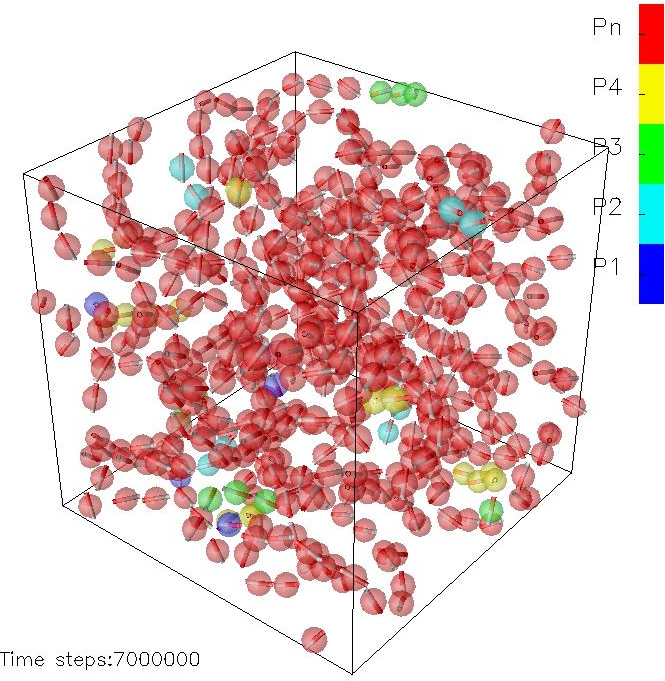} \\
(a) & (b)
\end{tabular}
\caption{\label{fig:three} Snapshots from LB simulations of $N=529$ 
colloids at a volume fraction $\phi=0.10$: (a) $\lambda=4$; (b) 
$\lambda=8$. Each particle is color-coded to reflect the total number of 
particles in the cluster to which it belongs: (dark blue) monomers; (light 
blue) dimers; (green) trimers; (yellow) tetramers; (red) clusters with 5 
or more particles.}
\end{figure}

Overall, the observed behavior of $g(r)$ and $h_{l_{1}l_{2}m}(r)$ is in 
good accord with that established in earlier studies 
\cite{Levesque:1994/a,Weis:1993/b}. With $\lambda=0$, $g(r)$ shows only 
short-range correlations (data not shown); with $\lambda \gg 1$, the 
primary peak in $g(r)$ becomes very strongly pronounced due to the high 
degree of particle association to form, at these low densities, 
chain-like aggregates. $h_{110}(r)$ helps distinguish parallel from 
anti-parallel correlations between the dipole moments, $h_{112}(r)$ 
contains (minus) the dipolar potential, and $h_{220}(r)$ picks out 
`nematic' orientational ordering. All of these functions show positive 
peaks at short range -- at intervals close to $2a$ -- confirming the 
prevalence of the `nose-to-tail' parallel conformations of nearby dipole 
moments within chains. As expected, the peaks become more pronounced 
with increasing $\lambda$. For a visual confirmation of the chaining, 
Figure \ref{fig:three} shows two snapshots from equilibrated LB 
simulations of $529$-particle systems with $\lambda=4$ and $\lambda=8$ 
at $\phi=0.10$. Each particle is color-coded to reflect the total number 
of particles in the cluster to which it belongs; monomers, clusters with 
$n=2$-$4$ particles, and clusters with $n \geq 5$ particles are given 
unique colors. (See section \ref{sec:clusterstats}.) The chain-like 
structural motif is clearly visible in both systems. Figures 
\ref{fig:one} and \ref{fig:two} show that interparticle correlations are 
short-ranged (compared to the box dimensions), and hence the 
thermodynamic properties should not show any pronounced finite-size 
effects. Nonetheless, the cluster network in Figure \ref{fig:three} 
appears to span the simulation cell, and so we might anticipate some 
finite-size effects in the long-time, long-wavelength dynamics. We have 
simulated the largest possible system sizes throughout.

%%% Figure 3 %%%

\clearpage

\section{Dynamic correlators in equilibrium} \label{dynamics}

We now present results for the intermediate scattering function $S(q,t)$, 
and its orientational analogs. As for the static structure, we restrict 
attention to the case $\phi = 0.10$ where we can combine complete 
equilibration with adequate statistical averaging. We examined relaxations 
in the Fourier components of the number density and magnetization density 
at wavevectors $\bm{q}$ commensurate with the periodic boundary 
conditions; to improve the statistics, we averaged the appropriate 
correlators at wavevectors of the same magnitude $q=|\bm{q}|$.

\subsection{Intermediate scattering function $S(q,t)$}

\begin{figure}[htbp]
\centering
\begin{tabular}{cc}
\includegraphics[width=2in]{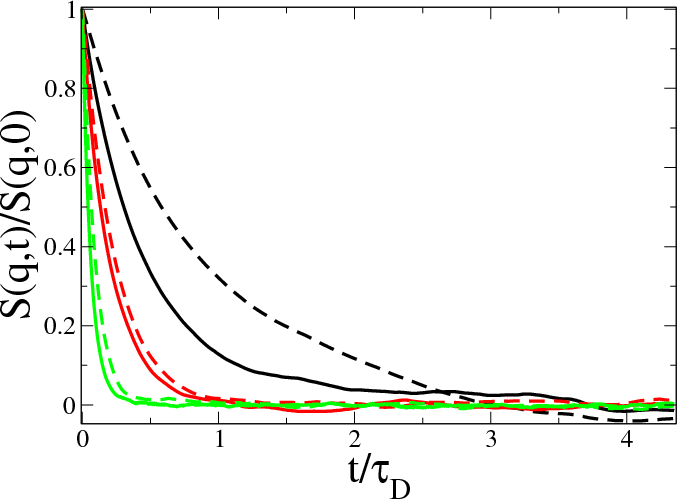} & 
\includegraphics[width=2in]{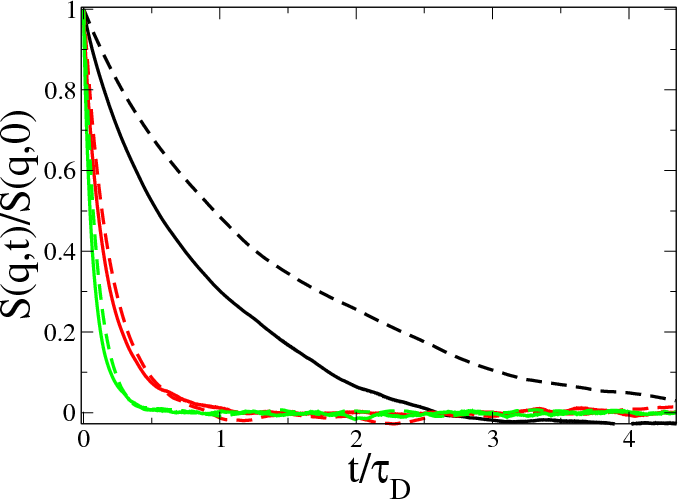} \\
(a) & (b) \\
\includegraphics[width=2in]{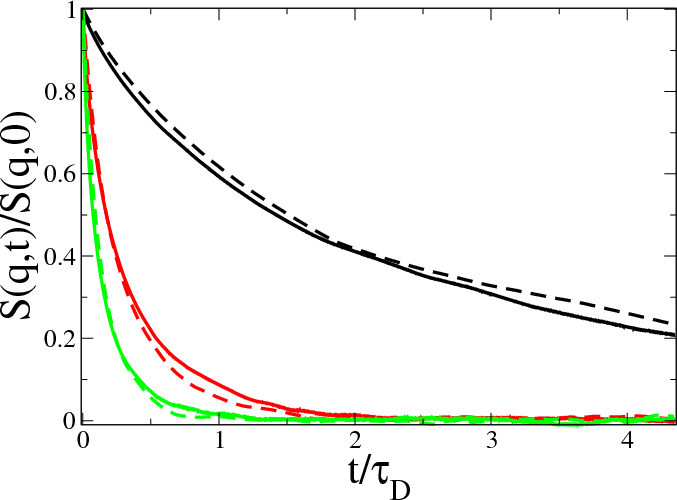} &
\includegraphics[width=2in]{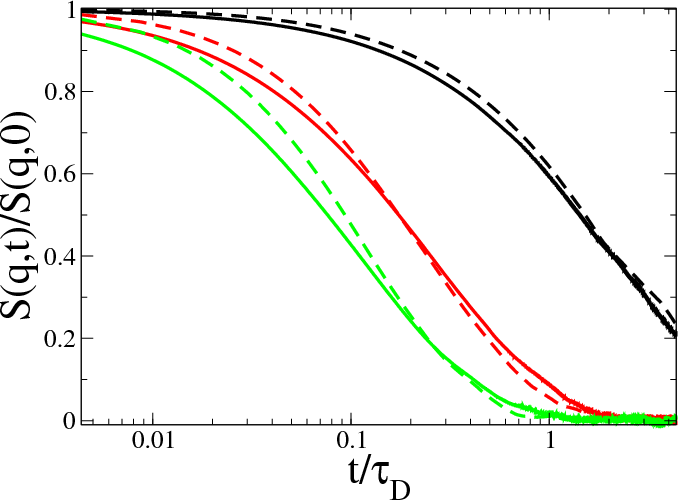} \\
(c) & (d)
\end{tabular}
\caption{\label{fig:four} Intermediate scattering functions with (a) 
$\lambda=0$, (b) $\lambda=4$, and (c) and (d) $\lambda=8$ with linear and 
logarithmic abscissas, respectively: (solid lines) BD; (dashed lines) LB. 
In each case the black lines are for $qa=1.1514$ and the green lines are 
for $qa=4.0456$. The red lines are for a wavevector close to the peak of 
$S(q)$, as follows: (a) $qa=2.6139$ at $\lambda=0$; (b) $qa=3.0715$ at 
$\lambda=4$; (c) and (d) $qa=3.2409$ at $\lambda=8$.}
\end{figure}

Figure \ref{fig:four} shows $S(q,t)$ (eq \ref{eq:S}) as a function of 
$t/\tau_D$ for $\lambda = 0$, $4$, and $8$ at $\phi = 0.10$. In each case 
curves are plotted for three different wavevectors; one close to the peak 
($q=q^*$) in the static structure factor $S(q)$, one larger, and one 
smaller. These linear-linear plots allow one to see clearly the long-time 
relaxation of the structure. (The short time dynamics is considered in 
section \ref{Hsection} below.)

%%% Figure 4 %%%

In all cases, the effect of hydrodynamic interactions is to slow down 
the relaxation of $S(q,t)$; this effect is most marked for wavenumbers 
well below $q^*$ (which in any case relax more slowly than those at the 
peak). The long-time relaxations are not far from exponential in all 
cases, and in particular show no sign of decomposition into separate 
$\alpha$ and $\beta$ relaxation processes as expected in colloidal 
systems on approach to a glass transition \cite{Pusey:1991/a}. Absence 
of the latter is confirmed by plotting the time on a logarithmic scale, 
as presented for $\lambda = 8$ in Figure \ref{fig:four}(d). It is 
notable, however, that the slowing by hydrodynamic interactions of the 
long-time relaxation, at least for $qa \simeq 1$, is much diminished at 
strong dipolar interactions ($\lambda = 8$). This might be taken as 
evidence that structural rearrangement in this case is controlled mainly 
by the energetics of aggregate rearrangement (breaking and reformation 
of dipolar contacts), and is no longer limited by the rate at which 
solvent can flow around the evolving structure -- a state of affairs 
generally expected to hold for glassy colloids. Caution is needed before 
drawing such a conclusion, however; many authors would, on adopting this 
reasoning, expect BD and LB curves to superpose only after rescaling of 
time by the short-time diffusion constant at the peak, $D_s(q^*)$ 
\cite{Fuchs:2005/a}. As discussed in section \ref{Hsection}, 
hydrodynamic interactions continue to cause a factor of 2 change in this 
quantity even for $\lambda = 8$.

\subsection{Orientational relaxation}

\begin{figure}[t!]
\centering
\begin{tabular}{cc}
\includegraphics[width=2in]{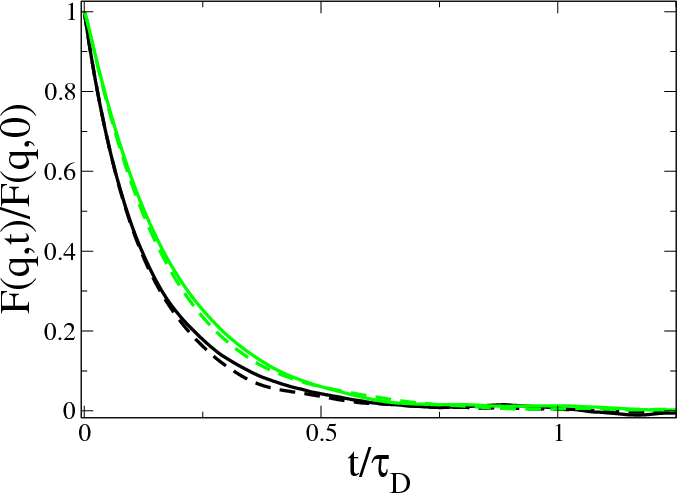} &
\includegraphics[width=2in]{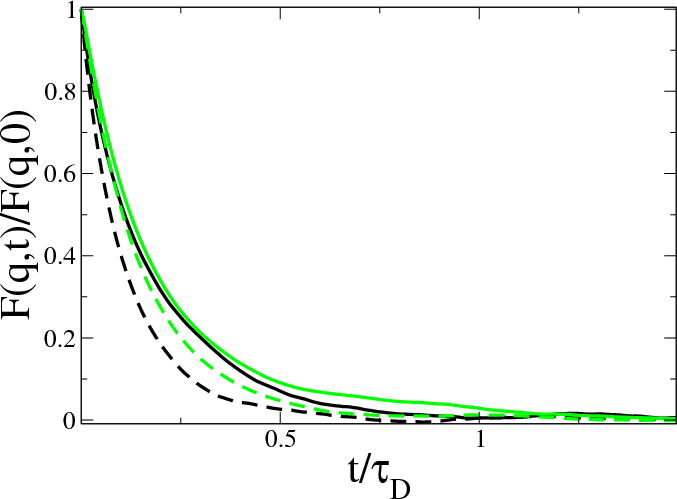} \\
(a) & (b) \\
\includegraphics[width=2in]{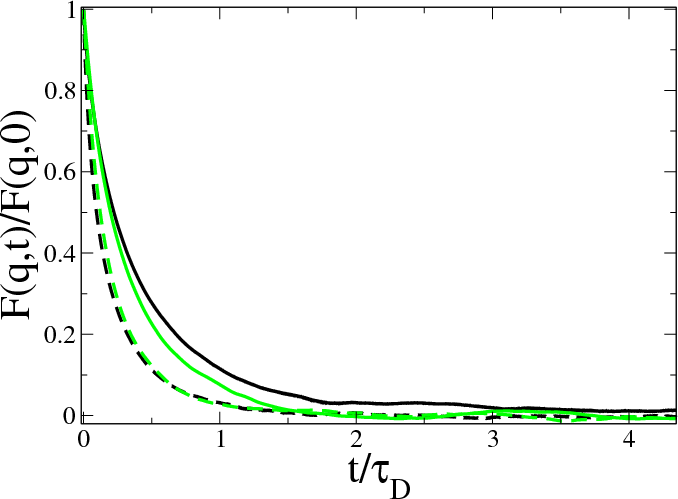} &
\includegraphics[width=2in]{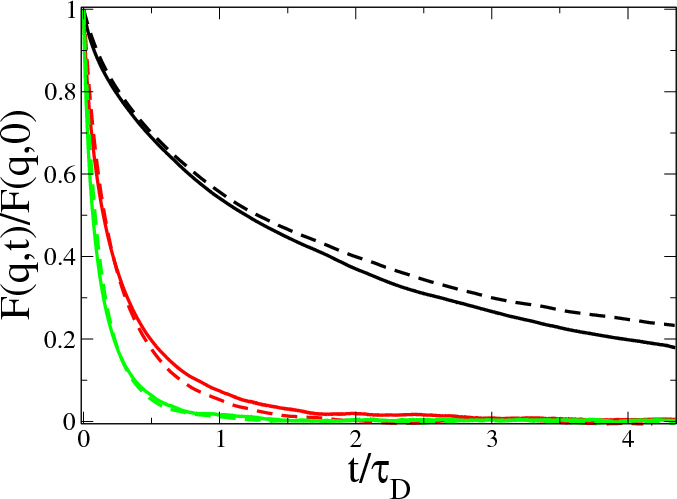} \\
(c) & (d)
\end{tabular}
\caption{\label{fig:five} Orientational relaxations resolved in to 
longitudinal ($F_{\rm L}$) and transverse ($F_{\rm T}$) correlation 
functions, at $\phi=0.10$ and (a) $\lambda=0$, (b) $\lambda=4$, and (c) 
and (d) $\lambda=8$. In (a), (b), and (c), black lines are BD and green 
lines are LB: (solid lines) $F_{\rm L}$; (dotted lines) $F_{\rm T}$. In 
(d), solid lines are BD and dashed lines are LB: (black lines -- upper) 
$qa=1.1514$; (red lines -- middle) $qa=3.2409$; (green lines -- lower) 
$qa=4.0456$.}
\end{figure}

Defining a wavevector-dependent dipole density $\bm{M}(\bm{q},t) = 
\sum_{j=1}^N\hat{\bm{s}}_j\exp{[-i\bm{q}\cdot\bm{r}_j(t)]}$ we can 
construct orientational correlators from the longitudinal (L) and 
transverse (T) components $\bm{M}_{\rm L} = 
(\bm{M}\cdot\hat{\bm{q}})\hat{\bm{q}}$ and $\bm{M}_{\rm T} = \bm{M} - 
\bm{M}_{\rm L}$ \cite{Felderhof:1993/a}:
\begin{eqnarray}
F(q,t) &=& N^{-1}
\langle\bm{M}(\bm{q},t)        \cdot\bm{M}(-\bm{q},0)         \rangle \\
F_{\rm L}(q,t) &=& N^{-1}
\langle\bm{M}_{\rm L}(\bm{q},t)\cdot\bm{M}_{\rm L}(-\bm{q},0) \rangle \\
F_{\rm T}(q,t) &=& N^{-1}
\langle\bm{M}_{\rm T}(\bm{q},t)\cdot\bm{M}_{\rm T}(-\bm{q},0) \rangle.
\end{eqnarray}
Data for $F_{\rm L}(q,t)$ and $F_{\rm T}(q,t)$ at $q\simeq q^*$ are 
plotted in Figure \ref{fig:five} for $\lambda = 0$, $4$, and $8$. For 
clarity, we omit $F(q,t)$ since this is a simple average of the 
longitudinal and transverse parts. The results are broadly comparable to 
the relaxation of $S(q,t)$ at similar wavevectors. We note that the 
longitudinal relaxations are slower than the transverse ones. This might 
be ascribed to the slow rotational diffusion of chain orientations with 
respect to the wavevector $\bm{q}$, as compared to faster librational 
motions of dipoles perpendicular to the local chain orientation. Figure 
\ref{fig:five}(d) shows (for $\lambda = 8$) the $q$ dependence of 
$F(q,t)$. This is again comparable to that for the density relaxation. 
Note, though, that $\bm{M}$ is not a conserved quantity and therefore, 
unlike the density, is not compelled to relax slowly for $qa\le 1$. The 
fact that it does so suggests that $\bm{M}$ is enslaved to slow particle 
rearrangements, as would arise if the dipole moments inside a cluster 
were to adopt frozen orientations relative to the positions of the 
constituent particles over the cluster's lifetime. With a dipolar 
bonding energy of $\simeq 16k_BT$ for two linearly aligned dipoles, such 
behavior is quite plausible.

%%% Figure 5 %%%

\subsection{Short time diffusion} \label{Hsection}

The shape of $S(q,t)$ is partly characterized by a short-time collective 
diffusion constant
\begin{equation}
D_s(q) = -\frac{1}{q^2}\left[\frac{d\ln S(q,t)}{dt}\right]_{s}
\label{D_s}
\end{equation}
where $[...]_s$ denotes a measurement taken at time scales long enough 
that a single particle indeed moves diffusively, but short enough that its 
average displacement remains small compared to $a$ (or, if smaller, the 
surface-to-surface separation from neighboring particles). For an isolated 
particle we therefore require $\tau_{\eta},\tau_{v} \ll t \ll \tau_{D}$ 
where we recall that $\tau_{\eta} = a^2\rho/\eta$ is the time scale for 
steady fluid motion to be established at the particle scale, $\tau_{v} = 
m/6\pi\eta a$ is the velocity autocorrelation time of the particle (of 
mass $m$), and $\tau_{D} = a^2/D_{0}$ is the time for a particle to 
diffuse its own radius. For particles with caging or bonding at 
surface-to-surface separations $h$, the requirement $t \ll \tau_{D}$ is 
replaced by $t \ll \tau_{D}(h/a)^2$. Defining
\begin{equation}
D_s(q) = \frac{D_{0}H(q)}{S(q)} 
\label{BD}
\end{equation}
one finds that in the absence of hydrodynamic interactions the 
`hydrodynamic factor' $H(q)$ is always unity for all $q$, whereas 
experiments on, e.g., hard sphere colloids show values that are not only 
smaller but also $q$-dependent \cite{Banchio:2008/a,Segre:1995/a}. For 
example, in hard-sphere colloids, $0.2\le H(q)\le 0.6$ at $\phi \simeq 
0.3$ and $1\le qa\le 4$ \cite{Banchio:2008/a,Segre:1995/a}, while 
$H(q^*)\simeq 0.8$ at $\phi \simeq 0.10$.

The numerical evaluation of $D_s(q)$, and hence of $H(q)$, carries 
significant difficulties associated with finite size corrections 
\cite{Ladd:1990/a}. That is, the long range nature of the hydrodynamic 
interactions, in conjunction with periodic boundary conditions, makes 
$D_s(q,N)$ very slow to converge with system size $V$, or equivalently 
with particle number $N = \phi V / v_{0}$ at fixed $\phi$. For the case of 
hard spheres, at least, this can be brought under good control at a 
semi-empirical level by adopting the following correction 
\cite{Ladd:1990/a,Segre:1995/a}
\begin{equation}
\frac{D_s(q)}{D_{0}} 
= \frac{D_s(q,N)}{D_{0}} 
+ \left(\frac{\eta_\infty}{\eta}\right)
  \left[ 
   1.7601 \left( \frac{\phi}{N} \right)^{1/3} - \frac{\phi}{N} 
  \right]
\label{laddcorr}
\end{equation}
where $\eta_\infty$ is the so-called high-frequency viscosity of the 
suspension and $\eta$ is the solvent viscosity.

To evaluate eq \ref{laddcorr} we numerically measured the high frequency 
viscosities for fully equilibrated systems with the given repulsive 
short-range potential and dipolar long-range interaction, using the recipe 
by Ladd \cite{Ladd:1994/a}. This amounts to calculating the integrated 
stress-stress correlation in a time window that is long enough to relax 
fluid degrees of freedom but too short for the colloids to move 
significantly. For $\lambda=0$, $4$, and $8$ we found $\eta_\infty/\eta$ 
to be $1.0532$, $1.0717$, and $1.1687$, respectively. A similar procedure 
was used in \cite{Banchio:2008/a} for the case of colloids with long range 
coulombic repulsions. However, since eq \ref{laddcorr} was invented to 
account for the observed finite-size behavior of systems of hard spheres, 
its use in other systems remains empirically questionable.  Below we 
therefore present data both for $D_{s}(q,N)$ as actually measured and for 
$D_s(q)$ as estimated via eq \ref{laddcorr}, but use the latter value to 
calculate $H(q)$.

The above caveat applies particularly when long-range (e.g., dipolar) 
interactions are present. Arguably such interactions should create their 
own finite size corrections, somewhat akin to those from hydrodynamics. In 
this case one might expect that, even with hydrodynamics switched off, the 
measured $H(q)$ would show size-dependent deviations from unity. In the 
data reported below we indeed find $H(q)$ values significantly less than 
unity for BD at large $\lambda$; however, we know of no method to correct 
for this and make no attempt to do so.

\begin{figure}[t!]
\centering
\begin{tabular}{ccc}
\includegraphics[width=2in]{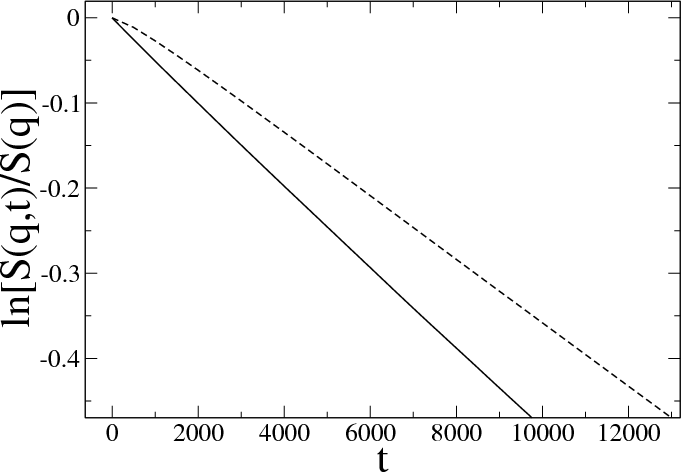} &
\includegraphics[width=2in]{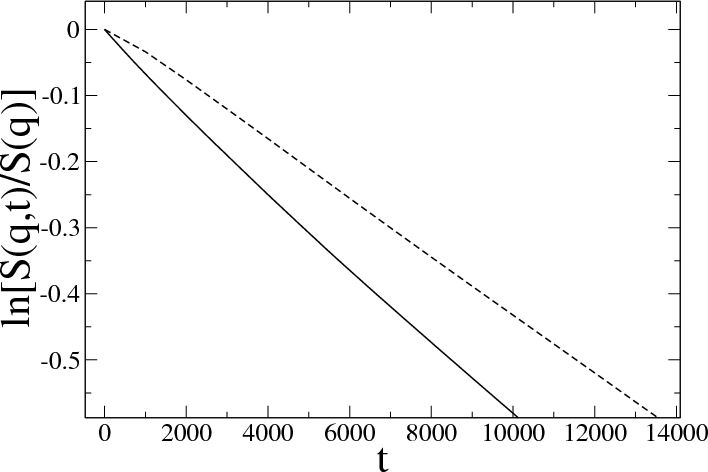} &
\includegraphics[width=2in]{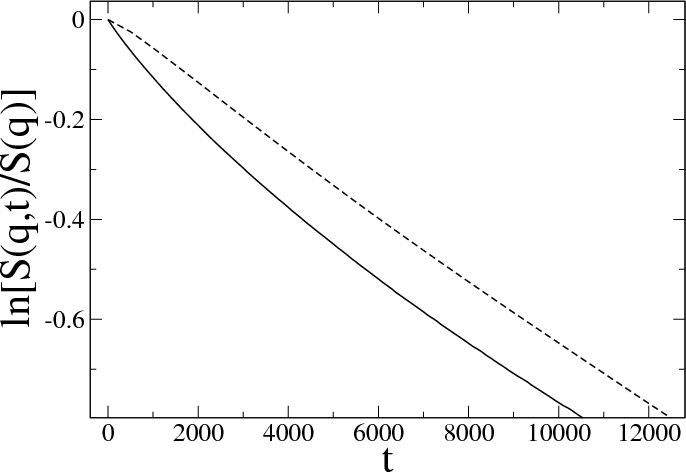} \\
(a) & (b) & (c) 
\end{tabular}
\caption{\label{fig:six} Short time decay of $\ln [S(q,t)/S(q)]$ as a 
function of $t$ for (a) $\lambda=0$, (b) $\lambda=4$, and (c) $\lambda=8$, 
showing the extent of the linear regime in each case. The solid lines are 
BD and the dashed lines are LB.}
\end{figure}

Figure \ref{fig:six} shows representative ($q\simeq q^*$) short time 
$S(q,t)$ data for the three values of $\lambda$ studied at $\phi = 0.10$. 
In accordance with expectation, the regime of short time diffusion is 
established beyond a few hundred timesteps, and for $\lambda = 0$ and $4$ 
there is thereafter a wide region of exponential decay within which 
$D_s(q)$ can be measured easily. For $\lambda = 8$ this window is 
foreshortened -- which is not surprising since the short time regime 
should end on the timescale of particle collisions. (For high interaction 
strengths, particles are bonded to neighbors with which they collide 
frequently.) Nonetheless a reasonable numerical estimate of the decay rate 
$D_s(q,N)$ can be made. In practice this was done by first identifying by 
eye the time window for short-time diffusion and then fitting to the 
log-linear plot within this window at each $q$. Finally the data for 
distinct $q$ values were binned (each bin containing roughly ten 
wavevectors) and the statistical error then estimated for the binned data.

%%% Figure 6 %%%

\begin{figure}[t!]
\centering
\begin{tabular}{cc}
\includegraphics[width=2in]{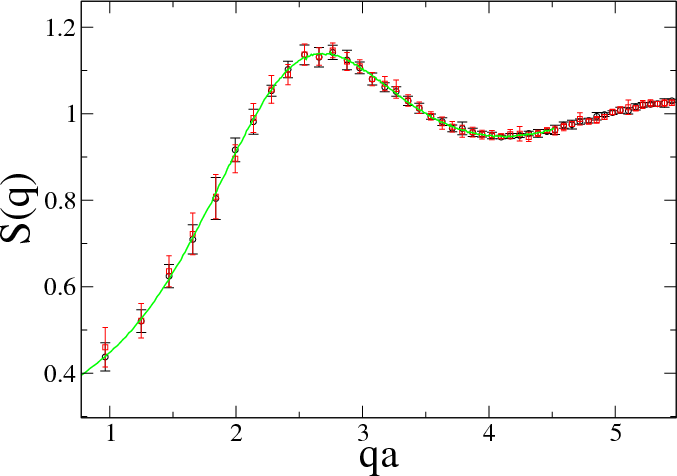} &
\includegraphics[width=2in]{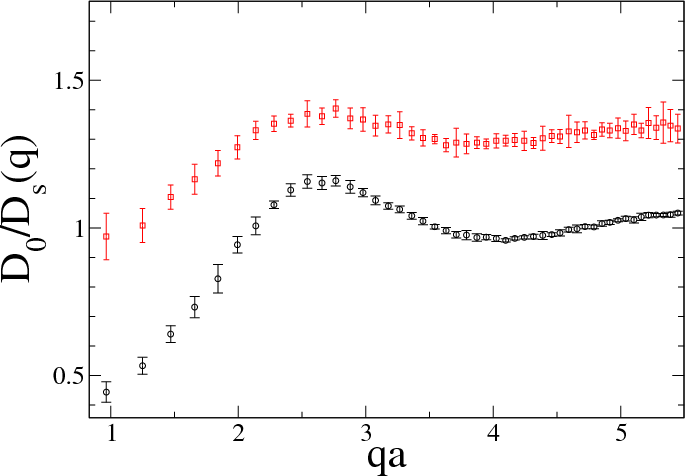} \\
(a) & (b) \\
\includegraphics[width=2in]{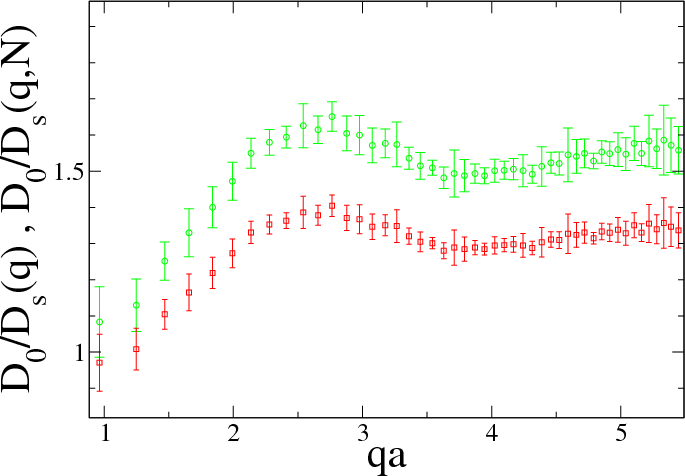} &
\includegraphics[width=2in]{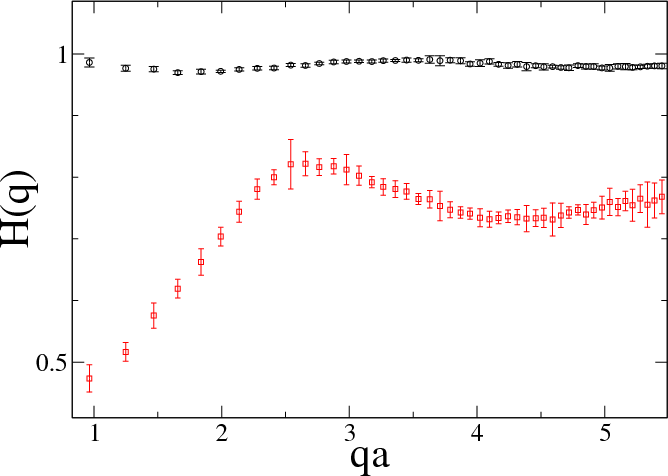} \\
(c) & (d)
\end{tabular}
\caption{\label{fig:seven} Structural and diffusion data for $\lambda=0$ 
and $\phi=0.10$: (black circles) BD; (red squares) LB; (green line) MC. In 
(c) the upper dataset (green) is with the uncorrected $D_s(q,N)$, and the 
lower dataset (red) is with the corrected $D_s(q)$.}
\end{figure}

\begin{figure}[htbp]
\centering
\begin{tabular}{cc}
\includegraphics[width=2in]{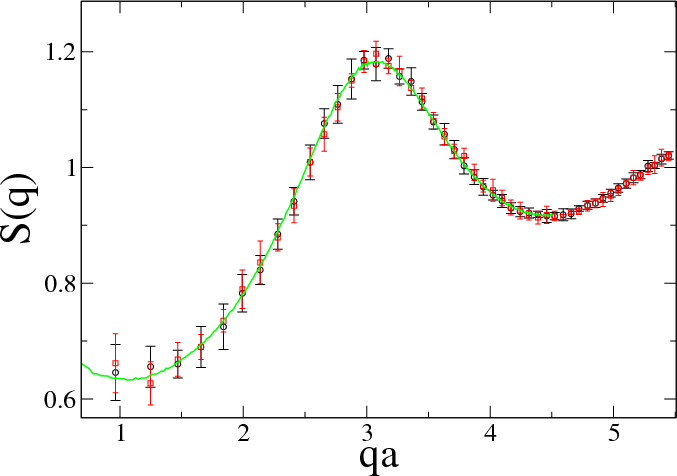} &
\includegraphics[width=2in]{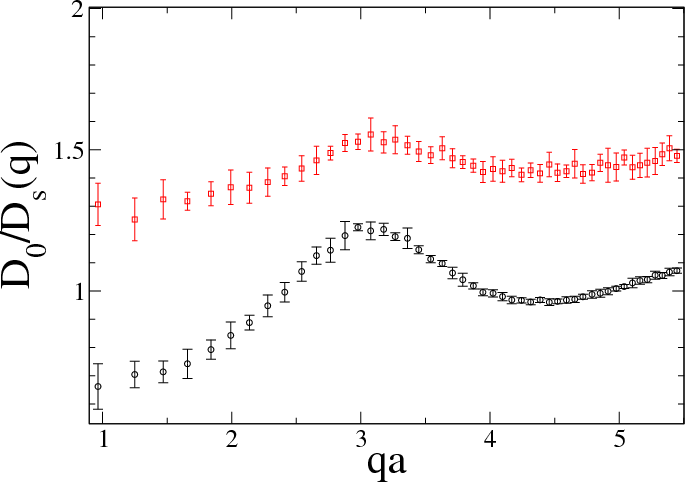} \\
(a) & (b) \\
\includegraphics[width=2in]{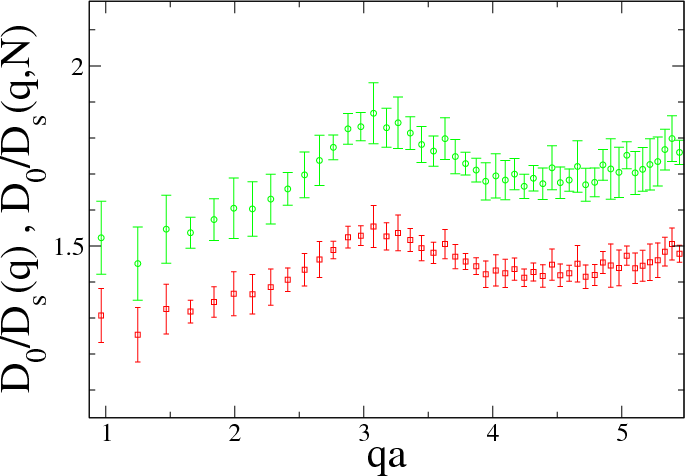} &
\includegraphics[width=2in]{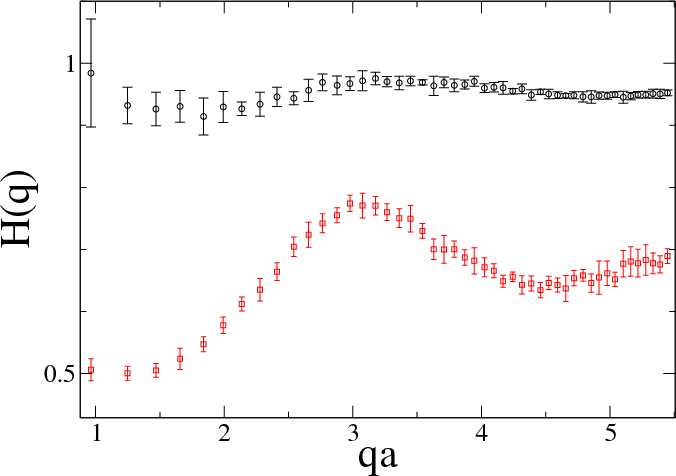} \\
(c) & (d)
\end{tabular}
\caption{\label{fig:eight} Structural and diffusion data for $\lambda=4$ 
and $\phi=0.10$: (black circles) BD; (red squares) LB; (green line) MC. In 
(c) the upper dataset (green) is with the uncorrected $D_s(q,N)$, and the 
lower dataset (red) is with the corrected $D_s(q)$.}
\end{figure}

\begin{figure}[htbp]
\centering
\begin{tabular}{cc}
\includegraphics[width=2in]{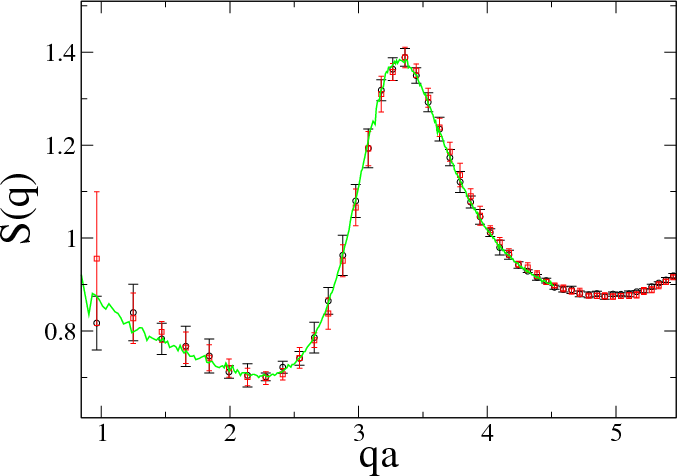} &
\includegraphics[width=2in]{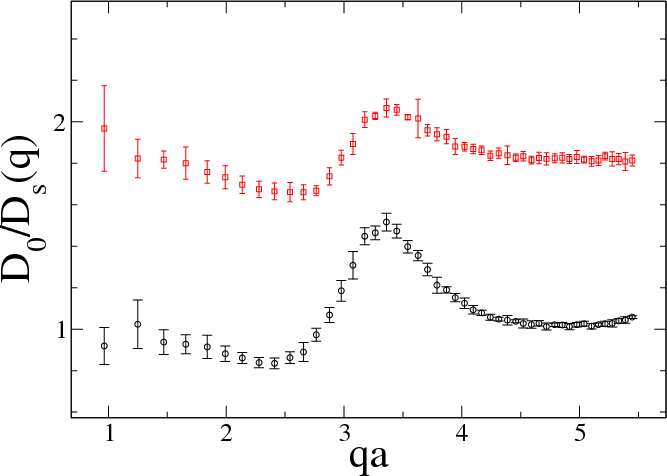} \\
(a) & (b) \\
\includegraphics[width=2in]{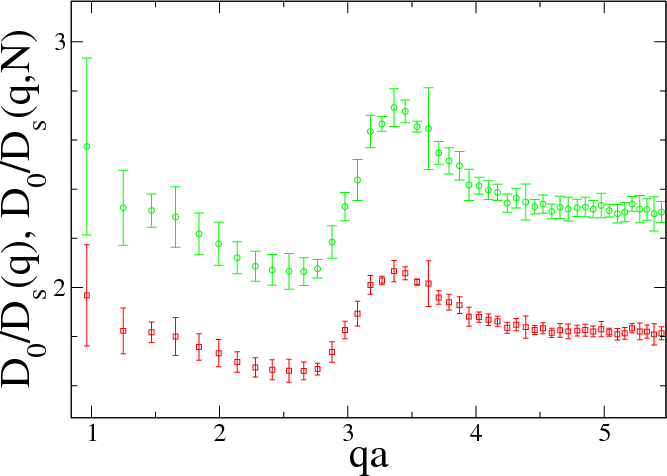} &
\includegraphics[width=2in]{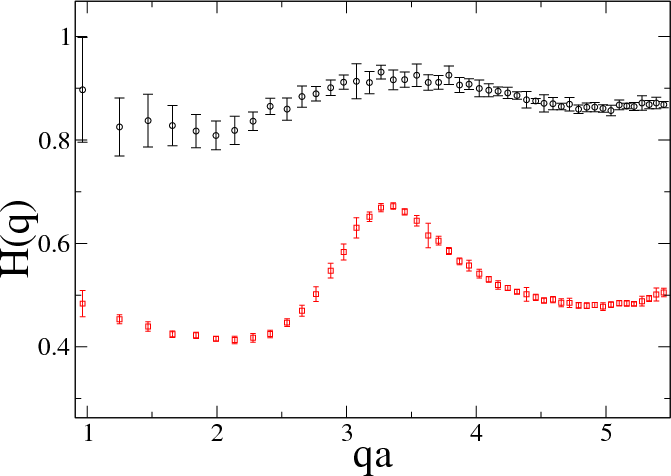} \\
(c) & (d)
\end{tabular}
\caption{\label{fig:nine} Structural and diffusion data for $\lambda=8$ 
and $\phi=0.10$: (black circles) BD; (red squares) LB; (green line) MC. In 
(c) the upper dataset (green) is with the uncorrected $D_s(q,N)$, and the 
lower dataset (red) is with the corrected $D_s(q)$.}
\end{figure}

Figures \ref{fig:seven}, \ref{fig:eight}, and \ref{fig:nine} show plots of 
$S(q)$, $D_0/D_s(q)$, and $H(q)$ generated from our dynamic datasets for 
$\lambda = 0$, $4$, and $8$, respectively. We also show, for comparison, 
the uncorrected $D_0/D_s(q,N)$ curves; $S(q)$ data generated from MC to 
check accuracy; and direct comparison with our BD results for $D_s(q)$ and 
$H(q)$. For the BD data no finite-size correction was made; for $\lambda = 
0$ we recover $H(q) = 1$ to simulation accuracy, with smaller values at 
larger $\lambda$ presumably attributable to finite size effects in the 
thermodynamic sector, as discussed above. (It is possible that, were these 
to be corrected, the $H(q)$ curves for LB could depend less strongly on 
$\lambda$ than in the results shown here.)

%%% Figure 7 %%%

%%% Figure 8 %%%

%%% Figure 9 %%%

When hydrodynamics is switched on, we obtain values of $H(q^*)\simeq 
0.6-0.8$ for all three $\lambda$ values that are comparable to previous 
data on hard sphere colloids at $\phi = 0.10$ 
\cite{Banchio:2008/a,Segre:1995/a}. However, for large $\lambda$, $S(q)$ 
and $H(q)$ both suggest a rising trend at small wavenumbers ($q\le 
q^*/3$). The rise in $S(q)$ at low $q$ is consistent with the formation of 
large dipolar clusters (and specifically, chains \cite{PJC:2000/c}). 
Clustering should also reduce the hydrodynamic friction per particle -- as 
is familiar from the fact that large clusters sediment more quickly under 
gravity. That is, the body force increases linearly with particle number 
$n$ whereas the viscous friction scales with hydrodynamic radius which is 
generally sublinear. This reduction is consistent with the observed rise 
in $H(q)$ at low $q$. Note however that this rise, although detectable 
beyond the scatter in the $H(q)$ data itself, is sensitive to the 
treatment of finite-size corrections and is therefore provisional.

\clearpage

\section{Transient dynamics and cluster formation} \label{kinetics}

We now consider the evolution of the structure following an initial 
quench, at time zero, from an equilibrium state with $\lambda = 0$ to one 
with either $\lambda = 4$ or $\lambda = 8$. We have studied such quenches 
at $\phi = 0.03, 0.10$, and $0.20$.

\subsection{Transient dipolar energy}

\begin{figure}[htbp]
\centering
\begin{tabular}{cc}
\includegraphics[width=3in]{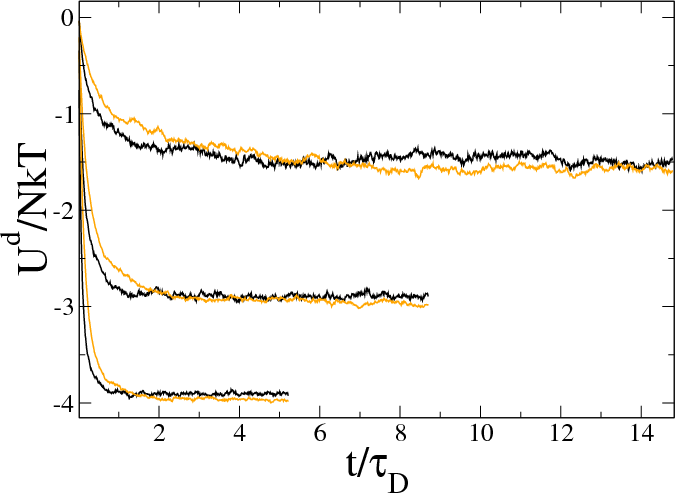} &
\includegraphics[width=3in]{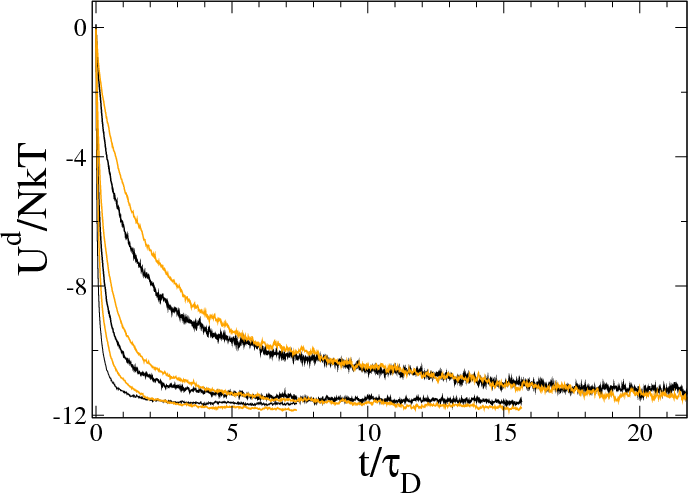} \\
(a) & (b)
\end{tabular}
\caption{\label{fig:ten} Relaxation of the dipolar energy following a 
quench from $\lambda = 0$ to (a) $\lambda = 4$ and (b) $\lambda = 8$: 
(black lines) BD; (orange lines) LB. Pairs of BD/LB curves correspond to 
the volume fractions, from top to bottom, $\phi=0.03, 0.10$, and $0.20$.}
\end{figure}

Figure \ref{fig:ten} shows the relaxation of the dipolar energy for each 
volume fraction as a function of time $t$ following the quench; LB and BD 
data are directly compared. In all cases, the effect of hydrodynamic 
interactions is to slow the approach to equilibrium. However, the effect 
is quite modest, and comparable to that reported earlier for $S(q,t)$ in 
equilibrium. The relaxation time is increased by no more than roughly a 
factor two, even for $\lambda = 8$. Note that addition of hydrodynamics is 
by no means guaranteed to slow down, rather than speed up, the approach to 
equilibrium. A familiar counterexample is binary fluid phase separation, 
where fluid flow of the two species creates a less dissipative, and hence 
faster, phase-separation route than pure diffusion at intermediate and 
late times \cite{Kendon:2001/a,Stansell:2006/a,Stratford:2007/a}.

%%% Figure 10 %%%

\subsection{Cluster statistics} \label{sec:clusterstats}

We define two dipolar particles to be in a bonded configuration if their 
pair dipolar interaction energy $U^d$ from eq \ref{dipoleU} obeys
\begin{equation}
U^d < -0.75\lambda.
\label{bonding}
\end{equation}
This definition is somewhat arbitrary: in principle once could choose 
either an energy-based or a geometric criterion. Our choice corresponds to 
an energy criterion set by an equipotential surface, in configuration 
space, of the dipolar part of the interaction. This is more suitable than 
a criterion based solely on $r$: the latter would count as a bond any 
close encounter between dipoles even if their orientation was such as to 
create a strongly repulsive force. In addition, our choice is designed to 
capture end-to-end bonding but reject most encounters between antiparallel 
dipoles even when their orientation is such as to create a bond. (Because 
of the short-range repulsion, the energy minimum for such bonds lies above 
the threshold in eq \ref{bonding}.) The particular value of the energy 
threshold -- which is intermediate between those used in earlier studies 
\cite{Stevens:1994/a,Tavares:1999/a} -- gave cluster distributions in good 
accord with what was expected from visual inspection of simulation 
snapshots, and was sufficient for the current purpose of examining 
transient cluster formation.

\begin{figure}[tbp]
\centering
\begin{tabular}{cc}
\includegraphics[width=2in]{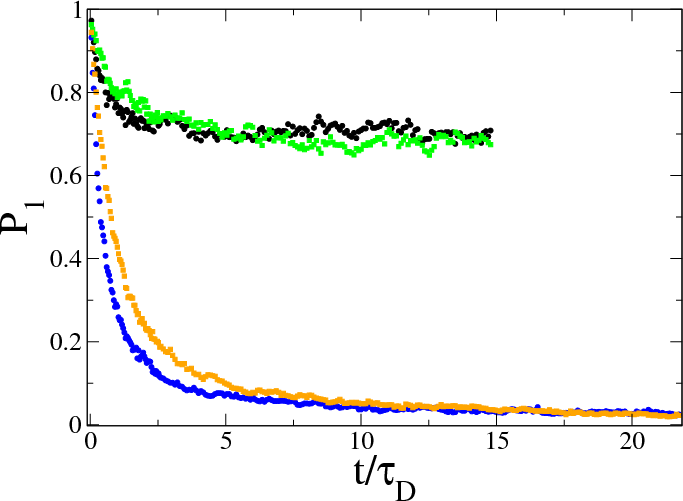} &
\includegraphics[width=2in]{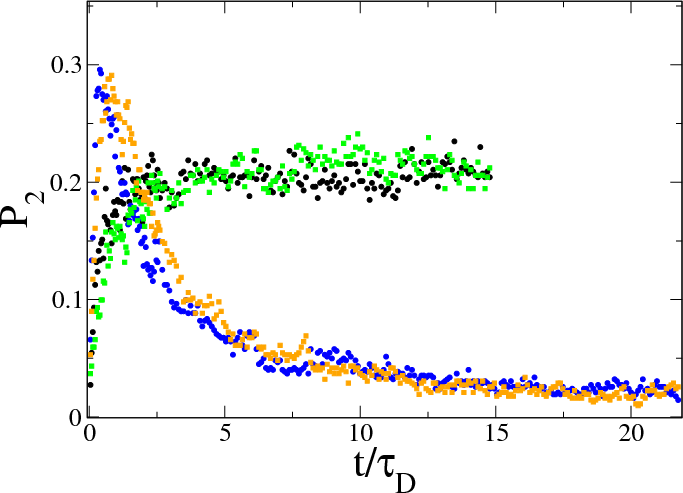} \\
(a) & (b) \\
\includegraphics[width=2in]{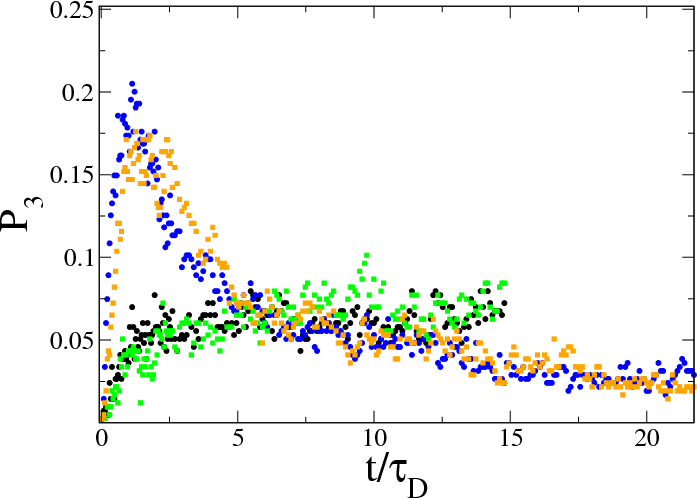} &
\includegraphics[width=2in]{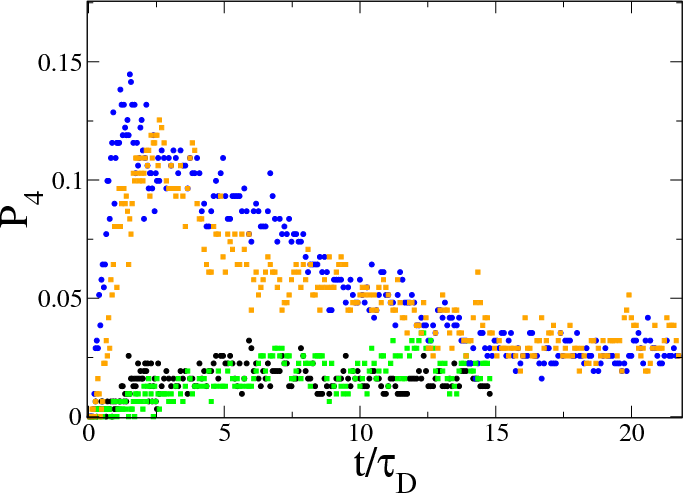} \\
(c) & (d)
\end{tabular}
\caption{\label{fig:eleven} Relaxation of cluster probabilities $P_n(t)$ 
following quenches from $\lambda=0$ to $\lambda = 4$ and $8$ at $\phi = 
0.03$: (black circles) BD with $\lambda=4$; (green squares) LB with 
$\lambda=4$; (blue circles) BD with $\lambda=8$; (orange squares) LB with 
$\lambda=8$.}
\end{figure}

\begin{figure}[htbp]
\centering
\begin{tabular}{cc}
\includegraphics[width=2in]{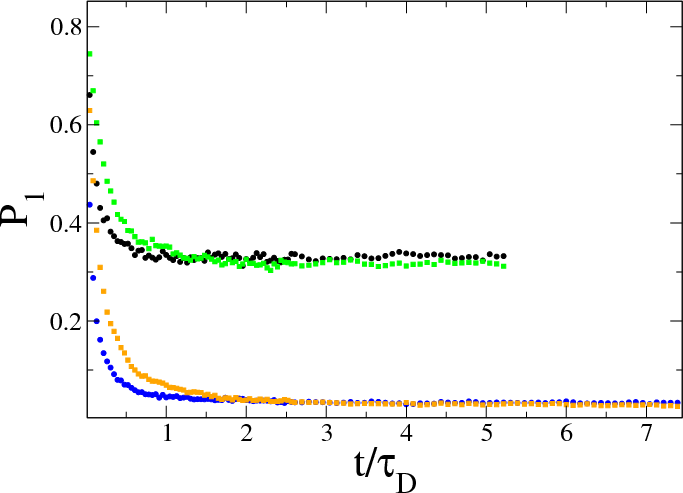} &
\includegraphics[width=2in]{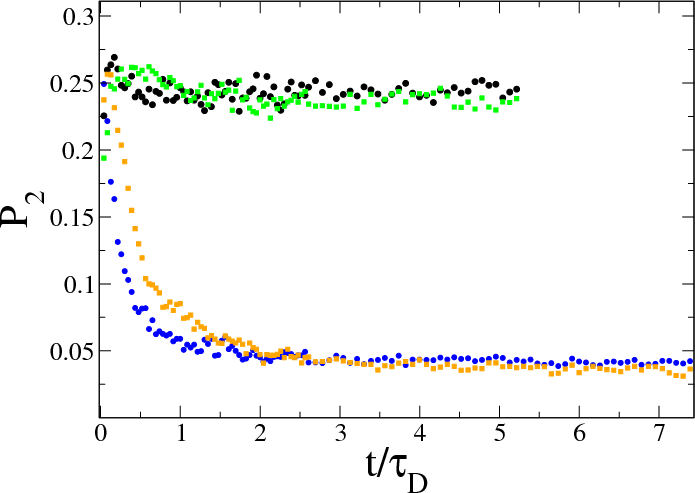} \\
(a) & (b) \\
\includegraphics[width=2in]{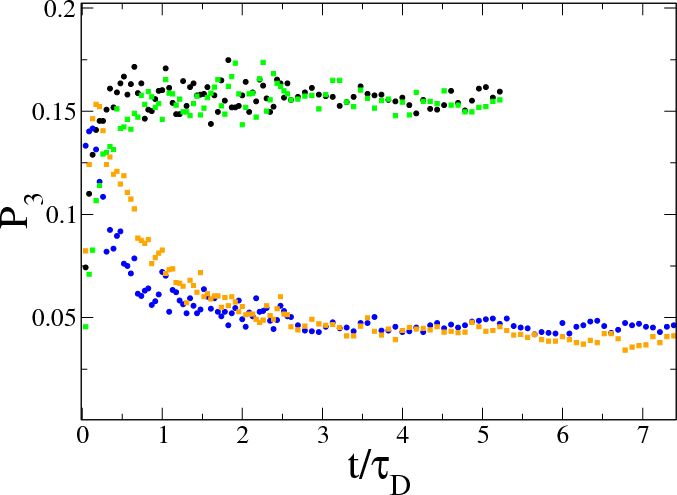} &
\includegraphics[width=2in]{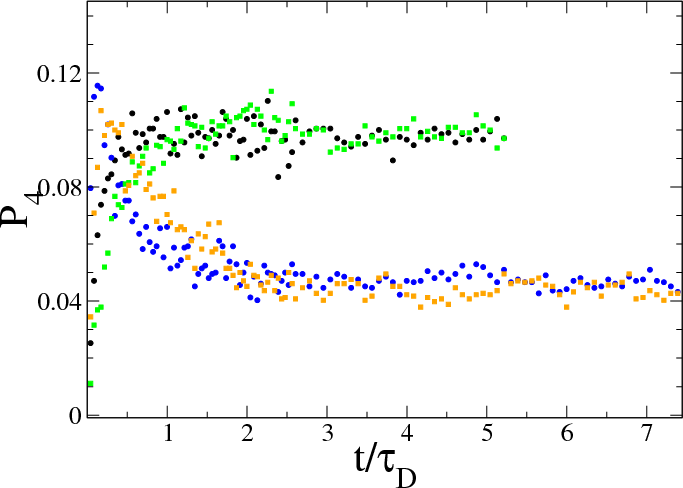} \\
(c) & (d)
\end{tabular}
\caption{\label{fig:twelve} Relaxation of cluster probabilities $P_n(t)$ 
following quenches from $\lambda=0$ to $\lambda = 4$ and $8$ at $\phi = 
0.20$. Symbols as in Figure \ref{fig:eleven}.}
\end{figure}

Using eq \ref{bonding} we partition each configuration of $N$ particles 
into a set of disjoint clusters, and monitor the fraction $P_n$ of 
particles that are assigned to clusters of size $n$. The time evolution 
of $P_n(t)$ gives information about the growth of clusters following the 
quench from $\lambda = 0$ at $t=0$. Figures \ref{fig:eleven} and 
\ref{fig:twelve} show $P_n(t)$ data for various $\lambda$ at volume 
fractions $\phi=0.03$ and $0.20$. Once again, BD data is included for 
comparison. (The data is binned timewise with a stride of 25 timesteps 
for $t<200,000$ and 50 timesteps thereafter. This choice offered the 
best compromise between smoothness and sensitivity. The actual numbers 
of clusters are low, and the relative fluctuations are high, so it is 
not easy to iron out the noise.) The transient $P_n(t)$ dynamics is 
subject to a similar slowing by hydrodynamic interactions as was the 
energy transient. Other than this there are no obvious differences 
between the LB and BD data. For large $\lambda$, both show 
characteristically peaked plots for $P_{2}$, $P_{3}$, and $P_{4}$ as 
small clusters build up and are then subsumed into larger ones.

%%% Figure 11 %%%

%%% Figure 12 %%%

\begin{figure}[htbp]
\centering
\begin{tabular}{cc}
\includegraphics[width=3in]{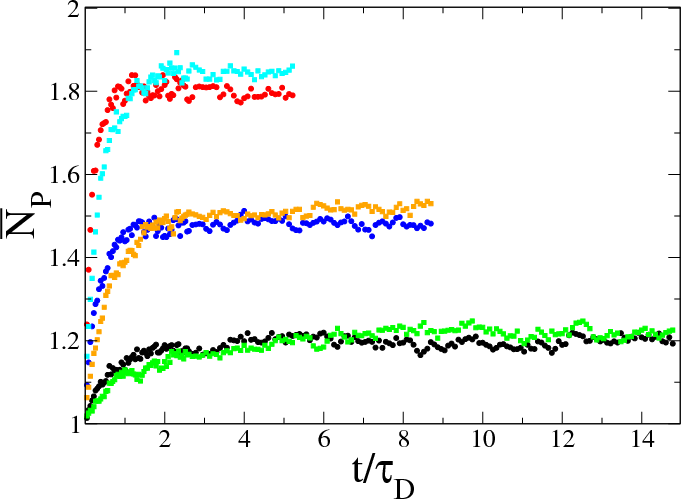} &
\includegraphics[width=3in]{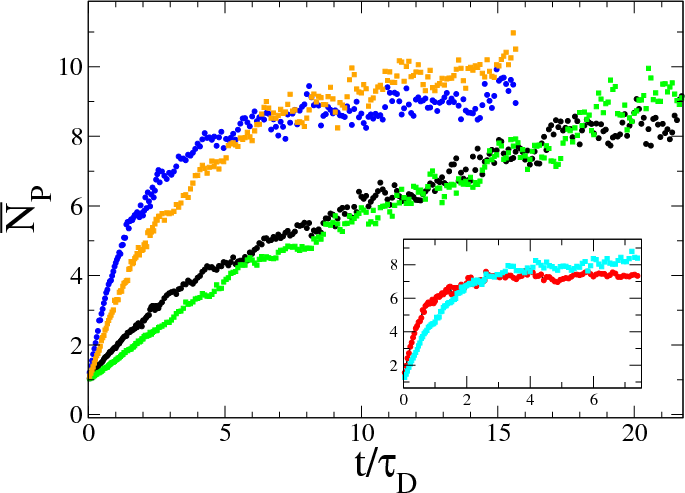} \\
(a) & (b)
\end{tabular}
\caption{\label{fig:thirteen} Time evolution of mean cluster size for (a) 
$\lambda = 4$ and (b) $\lambda = 8$, at $\phi = 0.03, 0.10, 0.20$: (black 
circles) BD with $\phi=0.03$; (green squares) LB with $\phi=0.03$; (blue 
circles) BD with $\phi=0.10$; (orange squares) LB with $\phi=0.10$; (red 
circles) BD with $\phi=0.20$; (cyan squares) LB with $\phi=0.20$.}
\end{figure}

Finally, in Figure \ref{fig:thirteen} we show the mean number of 
particles per cluster, $\bar{N}_{p}$, as a function of time for the same 
runs. The same hydrodynamic slowing is evident. Mean cluster sizes in 
the LB simulations are slightly higher than those in the BD simulations. 
This can be traced back to the small errors in dealing with particles 
close to contact, leading to more pronounced near-neighbour 
correlations, as discussed in section \ref{statics}.

%%% Figure 13 %%%

\section{Summary and Conclusions} \label{conclusions}

In this paper we have presented results for the equilibrium and transient 
dynamics of dipolar colloids with many-body hydrodynamic interactions. 
These were gained by incorporating the Ewald summation for the long-range 
dipolar interactions into our existing lattice Boltzmann algorithms, which 
handle the hydrodynamic forces by explicit propagation of momentum across 
the fluid residing on a lattice. The colloidal particles themselves are 
off-lattice and undergo molecular dynamics; Brownian motion is caused by 
fluctuating momentum transfer from the surrounding solvent, which creates 
correlated noise of the kind demanded by the fluctuation-dissipation 
theorem for hydrodynamically interacting particles. The full, fluid-driven 
noise can easily be replaced by local noise with no correlation between 
particles, creating a BD code. The resulting comparison of the LB and BD 
results allows the effect of many-body hydrodynamics to be isolated.

At the volume fractions ($0.03\le \phi \le 0.20$) and interaction 
strengths ($\lambda = 4$, $8$) studied here, these effects are easily 
measurable but remain relatively modest. Quantitative shifts in both 
short-time and long-time diffusion were observed for wavevectors near and 
below the peak in the static structure factor. Likewise, we found shifts 
in transient relaxation rates for the cluster size distribution on 
approach to steady state following a quench from $\lambda=0$. In all 
cases, the system with hydrodynamic interactions relaxes more slowly than 
the equivalent BD system. However, for the range of volume fraction and 
interaction strengths studied, this slowing down was rarely by more than a 
factor of two.

Although it is possible that stronger hydrodynamic effects would be 
observed in the dynamics of a quiescent system at larger $\phi$ and 
$\lambda$, their effects on long-time relaxation appear already to be 
decreasing for the largest values studied here. This could be a precursor 
to entering a glassy regime in which the crossing of local energy barriers 
limits relaxation rates; within this regime, conventional wisdom holds 
that hydrodynamics affects relaxational dynamics only through a scale 
factor \cite{Fuchs:2005/a}. However, despite the observation of slow 
transients and the difficulty of attaining full equilibration, even at 
$\phi = 0.20$ and $\lambda = 8$, we find no direct evidence for a glassy 
regime; specifically we see no separate $\alpha$ and $\beta$ relaxation 
processes. This is not surprising as our simulations run for at most 
$20-30\tau_D$, and a truly glassy system would certainly not approach 
equilibrium, as ours do, on this time scale.

Within our LB framework, there are serious obstacles to achieving much 
longer physical timescales using reasonable computational resources. One 
bottleneck remains the accurate treatment of noise; currently this 
requires a very large separation (order $10^5$ in our runs) between the 
simulation timestep and $\tau_D$. Future algorithmic work will, we hope, 
partially address this issue.

Our simulations were engineered to avoid the very large hydrodynamic 
forces that arise when hard colloidal particles come into lubrication 
contact. This was done by including a soft-core repulsion to maintain 
adequate separation between particle surfaces even when strongly bonded by 
the dipolar interactions. It is possible that these lubrication effects 
could enhance the relative role of hydrodynamic interactions, by further 
slowing the timescale for bond breakage and re-formation. To address this 
effect specifically (in a bulk periodic system) an algorithm such as ASD 
might be more suitable than LB. Note that within LB one can include a 
routine to address lubrication forces via an SD-like algorithm, but the 
computational scaling becomes bad when there are large clusters of 
particles in mutual lubrication contact. We do not know how well ASD would 
perform under such conditions, as compared to the purely repulsive 
interactions studied in \cite{Banchio:2008/a}.

Even without lubrication, the effects of many-body hydrodynamics could, of 
course, also become much more pronounced in various nonequilibrium 
situations. These include the rheological response to steady and/or 
time-dependent shearing, and perhaps the nonlinear response to large 
orienting fields. We hope to address one or more of these topics in future 
work.

\subsection*{Acknowledgments}

This work was funded in part under EPSRC Grants GR/S10377/01 and 
EP/C536452/1 (RealityGrid). We thank ECDF (Edinburgh Compute and Data 
Facility) for computational resources. EK thanks SUPA and ORS for a 
studentship. MEC holds a Royal Society Research Professorship.

\bibliographystyle{achemso}

\providecommand{\refin}[1]{\\ \textbf{Referenced in:} #1}

\end{document}